\documentclass[aps,prd,showpacs, superscriptaddress,  nofootinbib, floatfix, onecolumn,  amsmath,11pt, tightenlines]{revtex4}%

\usepackage{amssymb,hyperref, amsmath}
\usepackage[dvips]{color}
\usepackage[dvips]{graphicx}
\usepackage{epsfig}

\begin{document}

\preprint{JLAB-THY-09-949}

\title{Exotic and excited-state radiative transitions in charmonium from lattice QCD}

\author{Jozef J. Dudek}
\affiliation{Jefferson Laboratory, 12000 Jefferson Avenue,  Newport News, VA 23606, USA}
\email{dudek@jlab.org}
\affiliation{Department of Physics, Old Dominion University, Norfolk, VA 23529, USA}

\author{Robert G. Edwards}
\affiliation{Jefferson Laboratory, 12000 Jefferson Avenue, Newport News, VA 23606, USA}

\author{Christopher E. Thomas}
\affiliation{Jefferson Laboratory, 12000 Jefferson Avenue,  Newport News, VA 23606, USA}

\collaboration{for the Hadron Spectrum Collaboration }

\begin{abstract}

We compute, for the first time using lattice QCD methods, radiative transition rates involving excited charmonium states, states of high spin and exotics. Utilizing a large basis of interpolating fields we are able to project out various excited state contributions to three-point correlators computed on quenched anisotropic lattices. In the first lattice QCD calculation of the exotic $1^{-+}$ $\eta_{c1}$ radiative decay, we find a large partial width $\Gamma(\eta_{c1} \to J/\psi \gamma) \sim 100 \,\mathrm{keV}$.  We find clear signals for electric dipole and magnetic quadrupole transition form factors in $\chi_{c2} \to J/\psi \gamma$, calculated for the first time in this framework, and study transitions involving excited $\psi$ and $\chi_{c1,2}$ states.  We calculate hindered magnetic dipole transition widths without the sensitivity to assumptions made in model studies and find statistically significant signals, including a non-exotic vector hybrid candidate $Y_{\mathrm{hyb?}} \to \eta_c \gamma$. As well as comparison to experimental data, we discuss in some detail the phenomenology suggested by our results and the extent to which it mirrors that of quark potential models and make suggestions for the interpretation of our results involving exotic quantum numbered states.

\end{abstract}


\maketitle

\section{Introduction}
The charmonium system is often described as the ``hydrogen atom'' of meson spectroscopy.  Being reasonably non-relativistic, it is explained fairly well by potential models, at least below the open-charm ($D \bar{D}$) threshold.  More recently it has also been studied using effective field theory approaches (such as pNRQCD) and QCD sum rules.  There has lately been a resurgence of interest in the charmonium system with the $B$-factories, CLEO-c and BES finding missing states, making more accurate measurements of properties of these states and discovering a number of new resonances that are not easily explained by the quark model.  This has spurred renewed theoretical interest with much speculation as to whether these states are hybrids or multiquark/molecular mesons.  To date there are no charmonium states having manifestly exotic $J^{PC}$ such as $1^{-+}, 0^{+-}, 2^{+-}$, that would directly signal physics not present in potential models.

The states below open-charm ($D \bar{D}$) threshold can not decay via an OZI allowed strong decay and so have reasonably narrow widths.  Their radiative transitions can therefore have significant branching ratios and are experimentally accessible.  
The transitions from and production of the, as yet unobserved, exotic $1^{-+}$ are particularly interesting.  A lattice calculation of transition form factors of excited charmonia is therefore timely and this is the first such study.  The corresponding excited charmonium spectrum was calculated in lattice QCD in Ref.\ \cite{Dudek:2007wv}.  Transition form factors of the lightest few charmonia, those ground states accessible with interpolating fields $\bar{\psi}\Gamma \psi$, were calculated in Ref.\ \cite{Dudek:2006ej}; this work brought to the attention of CLEO-c experimentalists the discrepancy between the lattice calculated value of $\Gamma(J/\psi \to \eta_c \gamma)$ (and indeed the values predicted in most model calculations) and the single experimental measurement of this from Crystal Ball\cite{Gaiser:1985ix}. In a tour-de-force analysis\cite{Mitchell:2008fb}, a much more reliable value was extracted from CLEO-c data that is in much better agreement with theoretical estimates.

The calculation we will present is performed in the quenched approximation, neglecting altogether the effect of light quark degrees of freedom. As such it is rather directly related to the simplest quark potential models in which charm quarks move in a static potential of assumed gluonic origin. Attempts have been made to add in the effects of light-quark loops to these models\cite{Eichten:1978tg, Eichten:1979ms, Barnes:2007xu}, in some cases finding that these effects can be large\cite{Li:2007xr}. We will address this possibility in light of our results.

A strong motivation for developing the lattice QCD techniques required to extract excited and exotic state radiative transition matrix elements is the versatility of the method. We can use these methods, tested here in charmonium, at any computationally feasible quark mass. In particular this opens up the possibility of computing in a framework close to QCD the meson photocouplings that appear in the meson photo-production process to be utilized in the JLab 12 GeV GlueX experiment\cite{Meyer:2006az}. In this paper we will also compare results with the flux-tube model of gluonic excitations which to date is the only theoretical guide to the size of the hybrid couplings and hence production rates\cite{Close:2003fz, Close:2003ae}.

The paper is structured as follows: We begin in section \ref{tech} with a description of the technology used to construct three-point correlators and to project on to the contribution due to various excited states. In section \ref{res} we present our results for the transition form-factors between various meson states. A discussion of the phenomenology of these results in terms of experiment and models follows in section \ref{phensec} before we conclude in section \ref{conc}. Two appendices which consider some technical details complete the manuscript.

\section{Technology}\label{tech}
In this paper we will explore radiative transitions with charm-mass quarks using the quenched anisotropic lattices and clover fermion action described in \cite{Dudek:2007wv}. Radiative transition matrix-elements follow from vector current three-point functions, whose construction using sequential source technology is described in \cite{Dudek:2006ej}. As in \cite{Dudek:2006ej}, only connected diagrams are considered with the assumption that disconnected diagrams are negligible in comparison. The new item of technology in this study is the use of a large basis of meson interpolating fields as explored in two-point functions in \cite{Dudek:2007wv} with further interpretation in \cite{Dudek:2008sz}. 

These operators are lattice-discretised versions of the gauge-invariant fermion bilinears,
\begin{equation}
 	\bar{\psi}(x) \Gamma \overleftrightarrow{D}_i \overleftrightarrow{D}_j \ldots \psi(x), \nonumber
\end{equation}
where $\overleftrightarrow{D} = \overleftarrow{D} - \overrightarrow{D}$ is the gauge-covariant derivative operator. In order to improve the overlap on to lower lying mesons, the quark fields may also be smeared over space as $e^{\tfrac{1}{4}\sigma^2 D_i D_i}\psi(x)$. Linear combinations of these basic operators have been constructed which transform irreducibly under the lattice cubic symmetry and which have rather simple interpretations in the limit of zero lattice spacing, for details see \cite{Dudek:2007wv, Dudek:2008sz}. The operator set is large enough to have a considerable redundancy within a given quantum number sector, which can be utilized to extract excited states.

\subsection{``Ideal'' operators and eigenvector projection}

In \cite{Dudek:2007wv} the spectrum of charmonium was extracted by solution of the generalised eigenvalue problem\protect\footnote{repeated indices are summed}
\begin{align}
	C_{ij}(t) v^\mathfrak{n}_j &= \lambda_\mathfrak{n}(t) C_{ij}(t_0) v^\mathfrak{n}_j \label{gen_eig} \\
	\lambda_\mathfrak{n}(t) &\to e^{- E_\mathfrak{n} (t-t_0)} \nonumber \\
	v^{\mathfrak{n}\dag}_i C_{ij}(t_0) v^{\mathfrak{m}}_j &= \delta_{\mathfrak{nm}} \nonumber
\end{align}
which constitutes a best solution for the spectrum in the variational sense. The matrix of two-point correlators is constructed as $C_{ij}(t) = \langle 0 | \sum_{\vec{x}} {\cal O}_i(\vec{x},t) \, {\cal O}_j(\vec{0}, 0) | 0 \rangle$ where the sum over lattice sites forces the three-momentum of single particle states to be zero\footnote{In some cases we also considered non-zero three-momentum correlators - this will be discussed later}. This method relies upon a redundancy of operators ${\cal O}_i$ in any given quantum number sector.

Fitting the time-dependence of the principal eigenvalues, $\lambda_\mathfrak{n}(t)$, gives us the masses (or energies at finite three-momentum) of the states with the quantum numbers of the hermitian operators ${\cal O}_{i,j}$. Details of the two-point analysis, including the importance of the choice of timeslice $t_0$, are given in \cite{Dudek:2007wv}. The interpretation of the eigenvector $v^{\mathfrak{n}}$ in the solution of Eqn. \ref{gen_eig} can be expressed as follows: weighting the operators by this vector gives the optimal operator within the limited operator space for the state labeled by $\mathfrak{n}$. It is convenient to normalise these ``ideal'' operators in a manner which accounts for the value of $t_0$:  $\Omega^{\mathfrak{n}} = \sqrt{2 E_\mathfrak{n}} e^{- E_\mathfrak{n} t_0 /2 }  v^{\mathfrak{n}}_i {\cal O}_i$. 

Note that these eigenvectors are trivially related to the vacuum-operator-state matrix elements or ``overlaps'', $Z^\mathfrak{n}_i = \langle \mathfrak{n} | {\cal O}_i | 0 \rangle$, that appear in a bound-state spectral decomposition, 
\begin{equation}
	C_{ij}(t)  = \sum_\mathfrak{n} \frac{Z^{\mathfrak{n}*}_i Z^\mathfrak{n}_j }{2 E_\mathfrak{n}} e^{- E_\mathfrak{n}t}, \nonumber
\end{equation}
by $Z^\mathfrak{n}_j v_j^\mathfrak{m} = \sqrt{2 E_\mathfrak{n}} e^{E_\mathfrak{n} t_0/2} \delta_{\mathfrak{n,m}}$ or 
\begin{equation}
 Z^\mathfrak{n}_j = \sqrt{2 E_\mathfrak{n}} e^{E_\mathfrak{n} t_0/2}\, v^{\mathfrak{n}\dag}_i C_{ij}(t_0). \label{vZ}
\end{equation}

The procedure we shall follow in this paper is to compute three-point correlators having at the source (located on a fixed timeslice, $t_i$) a (smeared) local operator of the form $\bar{\psi}(\vec{0}, t_i) \Gamma \psi(\vec{0}, t_i)$ - all possible gamma matrices can be considered for the computing cost of a single ``forward'' propagator. In this work we consider $\Gamma = \gamma_5, \gamma_i, 1$ giving access to pseudoscalar, vector and scalar states at the source.

At the sink (located on a fixed timeslice, $t_f$) we use sequential-source technology using a broad selection of local and derivative-based operators as described in \cite{Dudek:2007wv} - for each operator at a given momentum (usually $\vec{p}_f=000$) we have the computing cost of a single ``backward'' propagator.  At the vector current insertion (inserted on all timeslices $t_i < t < t_f$)  we insert the local vector current with all possible lattice three-momenta $\vec{q}$ up to $|\vec{q}|^2 = 4$. Translational invariance ensures momentum conservation, selecting the correct value of $\vec{p}_i$ out of the sum over all momenta produced by a local operator.
\begin{widetext}
\begin{equation}
 	C_{\Gamma \mu j}(\vec{p}_i, \vec{p}_f ; t_i, t, t_f) = \Big\langle 0 \Big| \sum_{\vec{z}} e^{-i \vec{p}_f \cdot \vec{z} }  {\cal O}_j(\vec{z}, t_f) \,\cdot \, \sum_{\vec{y}} e^{i \vec{q} \cdot \vec{y} } j_\mu(\vec{y}, t) \,\cdot \, \bar{\psi}(\vec{0}, t_i) \Gamma \psi(\vec{0}, t_i) \Big| 0 \Big\rangle \label{raw}
\end{equation}

Three-point correlators for ``ideal'' operators are constructed from the ``raw'' correlators, eqn \ref{raw} by projecting with the appropriate eigenvector\footnote{A somewhat different application of the same basic idea has recently been presented in \cite{Burch:2008qx}},
\begin{align}
 	C_{\Gamma \mu \mathfrak{n}}(\vec{p}_i, \vec{p}_f ; t_i, t, t_f) &= \Big\langle 0 \Big| \sum_{\vec{z}} e^{-i \vec{p}_f \cdot \vec{z} }  \Omega^{\mathfrak{n}}(\vec{z}, t_f) \,\cdot \, \sum_{\vec{y}}e^{i \vec{q} \cdot \vec{y} } j_\mu(\vec{y}, t) \,\cdot \, \bar{\psi}(\vec{0}, t_i) \Gamma \psi(\vec{0}, t_i) \Big| 0 \Big\rangle \nonumber \\
&= \sqrt{2 E_\mathfrak{n}} e^{- E_\mathfrak{n} t_0 /2 }  v^{\mathfrak{n}*}_j  C_{\Gamma \mu j}(\vec{p}_i, \vec{p}_f ; t_i, t, t_f).\nonumber
\end{align}
\end{widetext}

The net effect is to provide us with a three-point correlator that, while it has multiple states contributing through the local source operator $\bar{\psi}(\vec{0}, t_i) \Gamma \psi(\vec{0}, t_i)$, should have only a single state, $\mathfrak{n}$, contributing from the sink operator $\Omega^{\mathfrak{n}}$. In principle we could improve this further by using the full basis of operators at the source as well as the sink, but using sequential source technology this quickly becomes rather expensive\footnote{In addition analysis of these correlators would require detailed knowledge of the finite-momentum behaviour of the derivative-based operators - we will consider this in appendix A and find that it is not entirely trivial}. 

In fact we can be a little more precise about the statement that only a single state contributes at the sink - in considering the analysis of the two-point correlators we determined that $t_0$ can be considered to be the timeslice on which the correlator matrix (of dimension $N$) is saturated by $N$ states - closer to the operator we require more states to describe the correlator matrix \cite{Dudek:2007wv, Blossier:2008tx}. It follows that if we are $t_0$ timeslices away from the sink operator position $t_f$ we can be quite confident that we have contribution only from the single state labeled by $\mathfrak{n}$. When we come to fit the time-dependence of the three-point correlator we will not use timeslices any closer to the sink than $t=t_f - t_0$.

\begin{figure}[h]
 \begin{center}
 \includegraphics[width=11.5cm,bb=0 0 960 662]{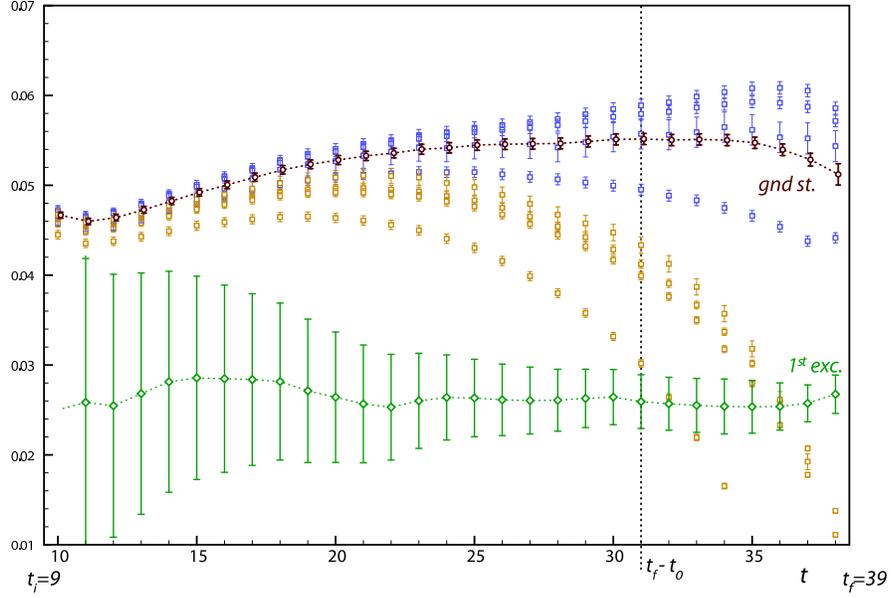}
\end{center}
\caption{\label{proj}Blue and yellow data - suitably normalised pseudoscalar-vector current-pseudoscalar correlators for a set of different sink operators. Red data - correlator formed by projection with the ground state eigenvector $v^{(0)}_i$. Green data - correlator formed by projection with the first excited state eigenvector $v^{(1)}_i$   }
\end{figure}

In figure \ref{proj} we show the effect of the eigenvector projection on to ``ideal'' states in the case of a pseudoscalar source operator $\bar{\psi}\gamma^5\psi$ and a selection of pseudoscalar sink operators. We clearly see that the projection on to the ground state ``ideal'' operator produces a considerable ``flattening'' of the correlator moving toward the sink. We do see some curvature beginning within $t_0 = 8$ timeslices away from the sink as expected\footnote{Indeed if we fit the time-dependence of the curvature at the sink with a single exponential we find that it corresponds to a mass heavier than the heaviest state extracted from the eight operator two-point correlation matrix}.  
More importantly for this study we see that projection on to the ``ideal'' operator for the \emph{first excited state} yields a clear non-zero signal from which one can extract a transition matrix element. 

\begin{figure}[h]
 \centering
 \includegraphics[width=11.5cm,bb=0 0 1100 642]{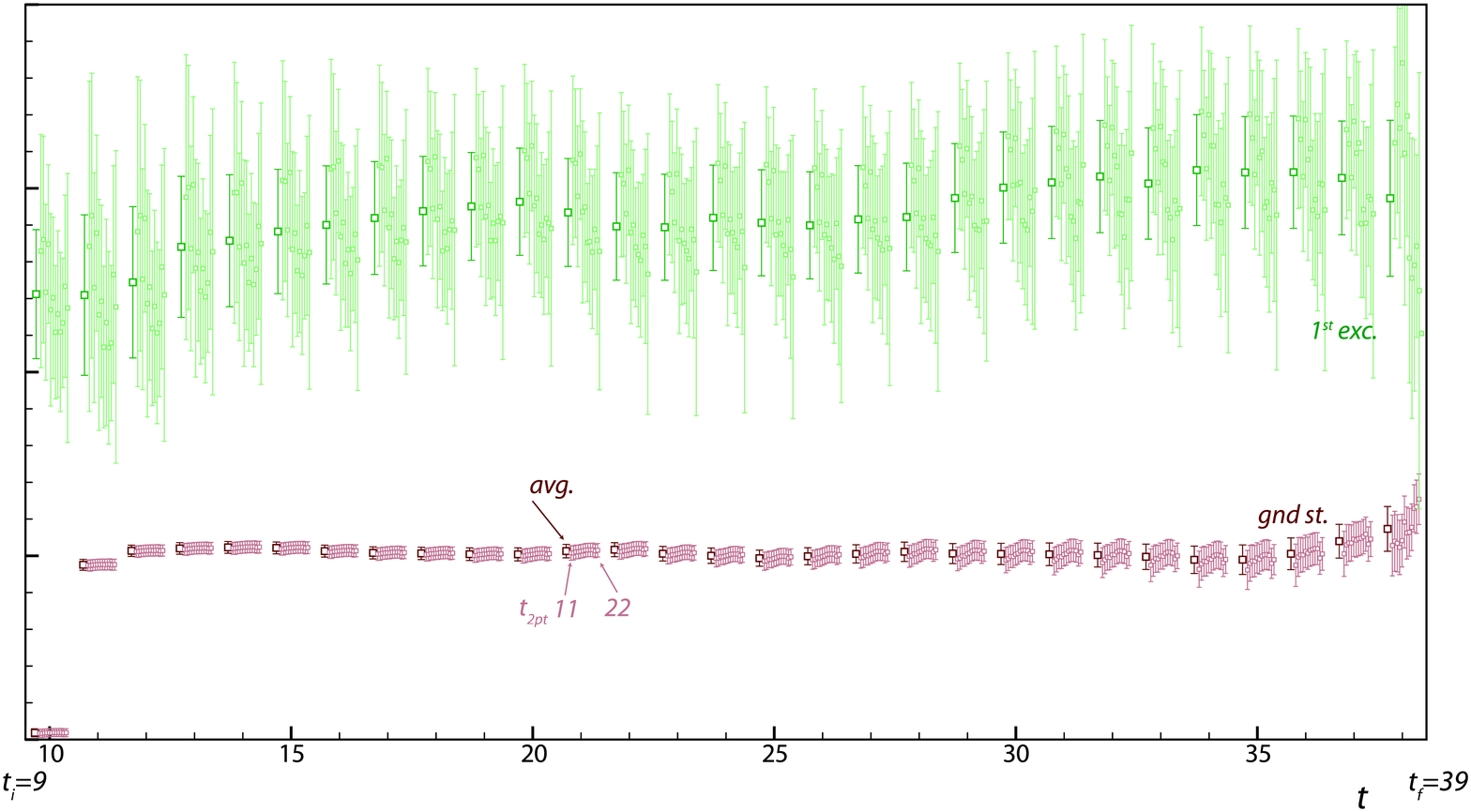}
\caption{\label{tZ}Eigenvector projected correlators as a function of current insertion time $t$. For each $t$ value we show the result of projecting with each of $v^{\mathfrak{n}=0,1}_i(t_\mathrm{2pt})$ for $11 \le t_\mathrm{2pt} \le 22$. Also shown is the average over the various $t_\mathrm{2pt}$ values. }
\end{figure}

There is in fact a slight technical subtlety to be dealt with in the eigenvector projection on to ``ideal'' operators that arises from the method of solution of the two-point problem. Solving the eigenvalue problem gives eigenvectors on each timeslice, $v^{\mathfrak{n}}_j(t_\mathrm{2pt})$ - if the solution is to be a true spectral representation there should be no time dependence in the eigenvectors, at least for $t_\mathrm{2pt} > t_0$ where the correlator matrix is supposed to be saturated. In \cite{Dudek:2007wv} we saw that over a considerable range of $t_\mathrm{2pt}$, the $Z$ values (obtained trivially from the $v$, see equation \ref{vZ}) were flat. We explore the effect of any such possible time-dependence on the three-point functions by performing the projection for eigenvectors belonging to a range of timeslices, $v^{\mathfrak{n}}_j(t_\mathrm{2pt})$. In figure \ref{tZ} we show for a ground state and first excited state projection the $t_\mathrm{2pt}$ dependence on the projected correlator. It is clear that there is no strong dependence on $t_\mathrm{2pt}$ and we choose to average over the projections for various $t_\mathrm{2pt}$ to reduce configuration-by-configuration fluctuations that may be caused by the generalized eigensystem solver.

\begin{widetext}
\subsection{Transition form-factors}
The ``ideal'' operator-projected three-point functions are related to the transition matrix elements between states as follows:
\begin{multline}
	C_{\Gamma \mu \mathfrak{n}}(\vec{p}_i, \vec{p}_f ; t_i, t, t_f) = \\ \Big\langle 0 \Big|  \Omega^{\mathfrak{n}}(\vec{0}, 0) \Big| f_{\mathfrak{n}}(\vec{p}_f) \Big\rangle   \frac{ e^{- E_{\mathfrak{n}} ( t_f - t)} }{2E_\mathfrak{n}}                     \sum_{\mathfrak{m}}   \Big\langle f_{\mathfrak{n}}(\vec{p}_f) \Big| j^\mu(\vec{0},0) \Big| i_{\mathfrak{m}}(\vec{p}_i)\Big\rangle \frac{e^{- E_{\mathfrak{m}} ( t - t_i) }}{2 E_\mathfrak{m}} \Big\langle i_{\mathfrak{m}}(\vec{p}_i) \Big| {\cal O}_i(\vec{0},0) \Big| 0 \Big\rangle \label{threepoint}
\end{multline}

While all possible eigenstates with the quantum numbers of ${\cal O}_i$ will contribute at the source (the sum over $\mathfrak{m}$), the eigenvector projection ensures that we need only consider the single state $\mathfrak{n}$ at the sink. Our normalisation of $\Omega_\mathfrak{n}$ ensures that if $f$ is spin-zero, $ \Big\langle 0 \Big|  \Omega^{\mathfrak{n}}(\vec{0}, 0) \Big| f_{\mathfrak{n}}(\vec{p}_f) \Big\rangle = 2 E_\mathfrak{n}$.

The Minkowski-space transition matrix element $ \big\langle f_{\mathfrak{n}}(\vec{p}_f) \big| j^\mu(\vec{0},0) \big| i_{\mathfrak{m}}(\vec{p}_i)\big\rangle$ can be decomposed in terms of multipole form-factors multiplied by Lorentz covariant combinations of the momenta and (if appropriate) polarization tensors of the particles labeled by $i,f$: $\sum_k   F_k(Q^2) \kappa_k^\mu( p_i, p_f, \epsilon_i, \epsilon_f^*)$. The general technique for obtaining these covariant multipole decompositions is given in the appendix of \cite{Dudek:2006ej}. Inserting this decomposition into eqn. \ref{threepoint}, and performing the implicit sums over helicity (in the case of spin $\ge 1$), the three-point correlator on each timeslice, $t$, can be expressed as a linear sum of known ``kinematic'' and ``propagation'' factors times the unknown multipole form-factors\footnote{This is essentially the same decomposition presented in \cite{Dudek:2006ej}}, 
\begin{equation}
 	C_{\Gamma \mu \mathfrak{n}}(\vec{p}_i, \vec{p}_f ; t_i, t, t_f) = \sum_\mathfrak{m} P_\mathfrak{m,n}(\vec{p}_i, \vec{p}_f, t_i, t, t_f) \sum_k K_{k,\mathfrak{m,n}}^\mu(\vec{p}_i, \vec{p}_f) F_{k,\mathfrak{m,n}}(Q^2). \nonumber
\end{equation}
In practice we opt to solve this linear system\footnote{by jackknifed SVD as described in \cite{Dudek:2006ej}} on each timeslice, $t$, for a set of ``effective'', $t$-dependent form-factors, $\tilde{F}_{k\mathfrak{n}}(Q^2,t)$, using the propagation and kinematic factors for the ground state at the source($\mathfrak{m} = 0$):
\begin{equation}
	 C_{\Gamma \mu \mathfrak{n}}(\vec{p}_i, \vec{p}_f ; t_i, t, t_f) = P_{0,\mathfrak{n}}(\vec{p}_i, \vec{p}_f, t_i, t, t_f) \sum_k K_{k,0,\mathfrak{n}}^\mu(\vec{p}_i, \vec{p}_f) \tilde{F}_{k,\mathfrak{n}}(Q^2, t). \nonumber
\end{equation}
\end{widetext}
The propagation factor $P_{0,\mathfrak{n}}(\vec{p}_i, \vec{p}_f, t_i, t, t_f)$ contains the overlap $Z_\Gamma^0(\vec{p}_i)$ and the energy $E_0(\vec{p}_i)$. These are determined in the analysis of two-point function at finite momentum. Since the state masses at zero momentum are rather precisely determined by the variational solution, we use the continuum dispersion relation to obtain $E(\vec{p}) = \sqrt{m^2 + |\vec{p}|^2}$. $Z$ values then follow from a linear fit to the finite-momentum two-point correlators supplying the known $\exp\left( -E_\mathfrak{n}(\vec{p}) t  \right)$ factors.

\begin{figure}
 \centering
 \includegraphics[width=10.5cm,bb=0 0 1079 663]{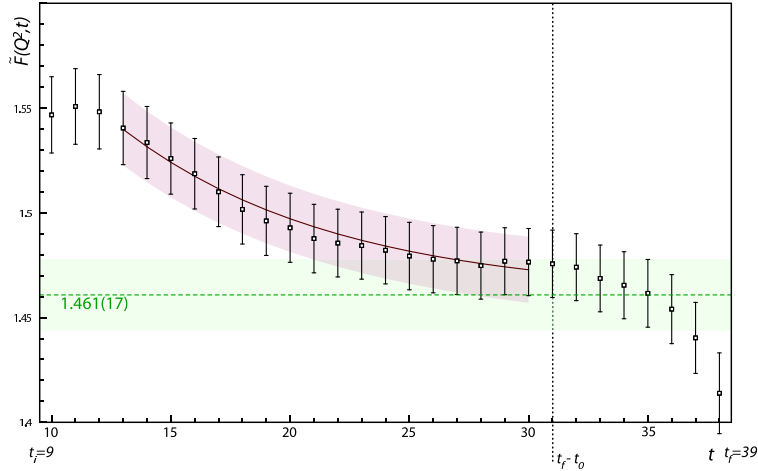}
\caption{\label{tdep}``Effective'' time-dependent form-factor. Fit with constant plus a single exponential shown in red, constant fit value shown in green. Fit is linear for the coefficients since the excited state energies appearing in the exponent are determined from solution of the two-point function problem.}
\end{figure}

The extracted ``effective'' form factors have the property that
$\tilde{F}_{k,\mathfrak{n}}(Q^2, t) \approx F_{k,\mathfrak{n}}(Q^2) +
{\cal O}\left(e^{- (E_{\mathfrak{m}=1} - E_{\mathfrak{m}=0}) (t -
    t_i)}\right)$ so that away from the source the excited state
contributions die away and the form-factor plateaus to the
$i_{\mathfrak{m}=0} \to f_{\mathfrak{n}}$ value. We fit the
time-dependence of these extracted ``effective'' form-factors to a sum
of exponentials where the energy dependence is that extracted from
spectrum studies\footnote{With the exponential dependence supplied
  this is a linear fit for the coefficients}. A typical example is
shown in figure \ref{tdep}. We retain only the plateau value, discarding the
relatively unreliable source excited state information.\footnote{Strictly speaking the kinematic factors for excited $\mathfrak{m}$ states need not be proportional to the ground state values such that excited state multipoles may ``leak'' into the wrong ground state multipole - clearly this effect will fall off for $t\gg t_i$. An alternative approach is to include the excited state terms directly into the decomposition fit, expanding the space to also be over timeslices, the time-dependence being the discriminator for the various states contributing at the source - this guarantees the right kinematic factors for excited states. In all cases to be presented the analysis was cross-checked using this fitting method (the excited state energies obtained from applying the dispersion relation to the excited state masses reliably extracted using the variational method) and agreement within statistical fluctuations found.}

The expansion of transition matrix elements as performed above relies upon the continuum Lorentz symmetry which is broken on our cubic lattice. A rigorous study would consider the expansion in terms of irreducible representations of the cubic group at finite momentum and would typically involve a larger number of form-factors, some of which would tend to the multipole form-factors as $a \to 0$ and some of which would have to vanish. We have not performed such a decomposition but suspect that it is not important at this lattice spacing since when we solve the correlator-form-factor linear system assuming continuum-like decompositions we typically obtain $\chi^2$ per d.o.f. very close to 1, suggesting that there is little need to enlarge the basis space. 

The fact that charmonium states are eigenstates of charge conjugation ($C$), coupled with the photon having $C = -1$, means that all charmonium radiation transitions are between states of opposite $C$. The origin of this is that the photon couples equally to the charm quark and the charm anti-quark. In our lattice calculation we need not impose this symmetry; by coupling only to the quark we can obtain transitions between states of equal $C$. One way of viewing this is that it is like having a $u,d$ pair but as heavy as the charm quark - this can useful for comparison with models and we will show results for a number of $C$-violating transitions.

\section{Results}\label{res}

\subsection{Vector current renormalization}\label{ZV}
Since we use the local vector current in the construction of our three-point functions we need to determine the renormalization constant, $Z_V$, to relate the extracted matrix elements to physical matrix elements. We do this by insisting that the pseudoscalar form-factor at zero $Q^2$ takes the value 1. On an anisotropic lattice we should allow there to be different renormalization constants for temporally and spatially directed currents, indeed we find $Z_{V(s)} = 1.23(2)$ and $Z_{V(t)} = 1.118(6)$. In all results presented below only the spatially directed current is used. See appendix B for a discussion of the effect of improvement of the vector current.

\subsection{Scalar - Vector transitions}
The first results we will present concern transitions between scalar ($0^{++}$) and vector ($1^{--}$) states of charmonium. Using a quark-smeared operator $\bar{\psi}\psi$ at the source and 13 vector operators\footnote{In the notation of \cite{Dudek:2007wv} they are the quark-smeared versions of $\gamma_i, \gamma_i \gamma_0, a_0\times \nabla_{T1}, a_1\times \nabla_{T1}, \rho\times \mathbb{D}_{T1}, \rho_{(2)}\times \mathbb{D}_{T1}, \pi \times \mathbb{B}_{T1},\pi_{(2)} \times \mathbb{B}_{T1}$ and the unsmeared versions of $\gamma_i, \gamma_i \gamma_0,  a_1\times \nabla_{T1},  \pi \times \mathbb{B}_{T1},\pi_{(2)} \times \mathbb{B}_{T1}$. In the $\vec{p}_f = (100)$ case the $\rho_{(2)} \times \mathbb{D}$ and $\pi_{(2)}\times \mathbb{B}$ operators were not included.  } at the sink we extracted transitions between the ground state scalar $\chi_{c0}$ and the lowest six vector states. In table \ref{vectorstates} we show the spectrum reported in \cite{Dudek:2008sz}, possible comparable experimental states and a model-dependent state assignment.

\begin{table}
 \begin{tabular}{c c c | c}
level & mass / MeV & suggested state  & model assignment\\
\hline
0 & $3106(2)$ & $J/\psi$ & $1\,^3S_1$\\
1 & $3746(18)$ & $\psi'(3686)$ & $2\,^3S_1$\\
2 & $3846(12)$ & $\psi_3$ & lat. artifact\\
3 & $3864(19)$ & $\psi''(3770)$ & $1\,^3D_1$\\ 
4 & $4283(77)$ & $\psi($`4040'$)$ & $3\,^3S_1$\\
5 & $4400(60)$ & $Y$ ? & hybrid
 \end{tabular} 
\caption{\label{vectorstates}$T_1^{--}$ spectrum extracted from two-point functions, suggested experimental state analogues and quark model bound-state assignments made in \protect \cite{Dudek:2008sz}. }
\end{table} 

This transition is characterized by two multipole amplitudes, a transverse electric dipole $E_1(Q^2)$ and a longitudinal $C_1(Q^2)$, the matrix element decomposition being
\begin{widetext}
\begin{align}
\langle S(\vec{p}_S) | j^\mu(0) | V(\vec{p}_V, \lambda) \rangle =\Omega^{-1}(Q^2) \Bigg( &E_1(Q^2) \Big[ \Omega(Q^2)  \epsilon^\mu(\vec{p}_V, \lambda) - \epsilon(\vec{p}_V, \lambda)\cdot p_S \big(  p_V^\mu p_V \cdot p_S - m_V^2 p_S^\mu \big) \Big] \nonumber\\  &\quad + \frac{C_1(Q^2)}{\sqrt{q^2}} m_V  \epsilon(\vec{p}_V, \lambda) \cdot p_S \Big[ p_V \cdot p_S (p_V+p_S)^\mu - m_S^2 p_V^\mu - m_V^2 p_S^\mu \Big] \Bigg).\label{SV}
\end{align}
\end{widetext}
where $\Omega(Q^2) = (p_V \cdot p_S)^2 - m_S^2 m_V^2$.

The ``vector'' operators used are, in fact, all in a particular irreducible representation of the cubic group at rest, namely $T_1$ which in the continuum limit contains spins $1,3,4 \dots$. By considering the degeneracy pattern and the values of the overlaps $Z^{(\mathfrak{n})}_j$, the two-point function analysis strongly suggests that the 2nd excited state in the $T_1^{--}$ sector is in fact a spin-3 state and as such we do not present results involving this state here\cite{Dudek:2007wv, Dudek:2008sz}. Additionally we do not show the amplitudes for the 4th excited state\footnote{In \cite{Dudek:2008sz} this state was identified with the quark model state $3\,^3S_1$} as the signals are statistically consistent with zero for all $Q^2$. Shown in figure \ref{SVfig} are the electric dipole transition form-factors for the four relevant states. See appendix A for a discussion of the inclusion of $\vec{p}_f = (100)$ correlators in this analysis.

\begin{figure}
 \includegraphics[width=8cm,bb=0 0 681 480]{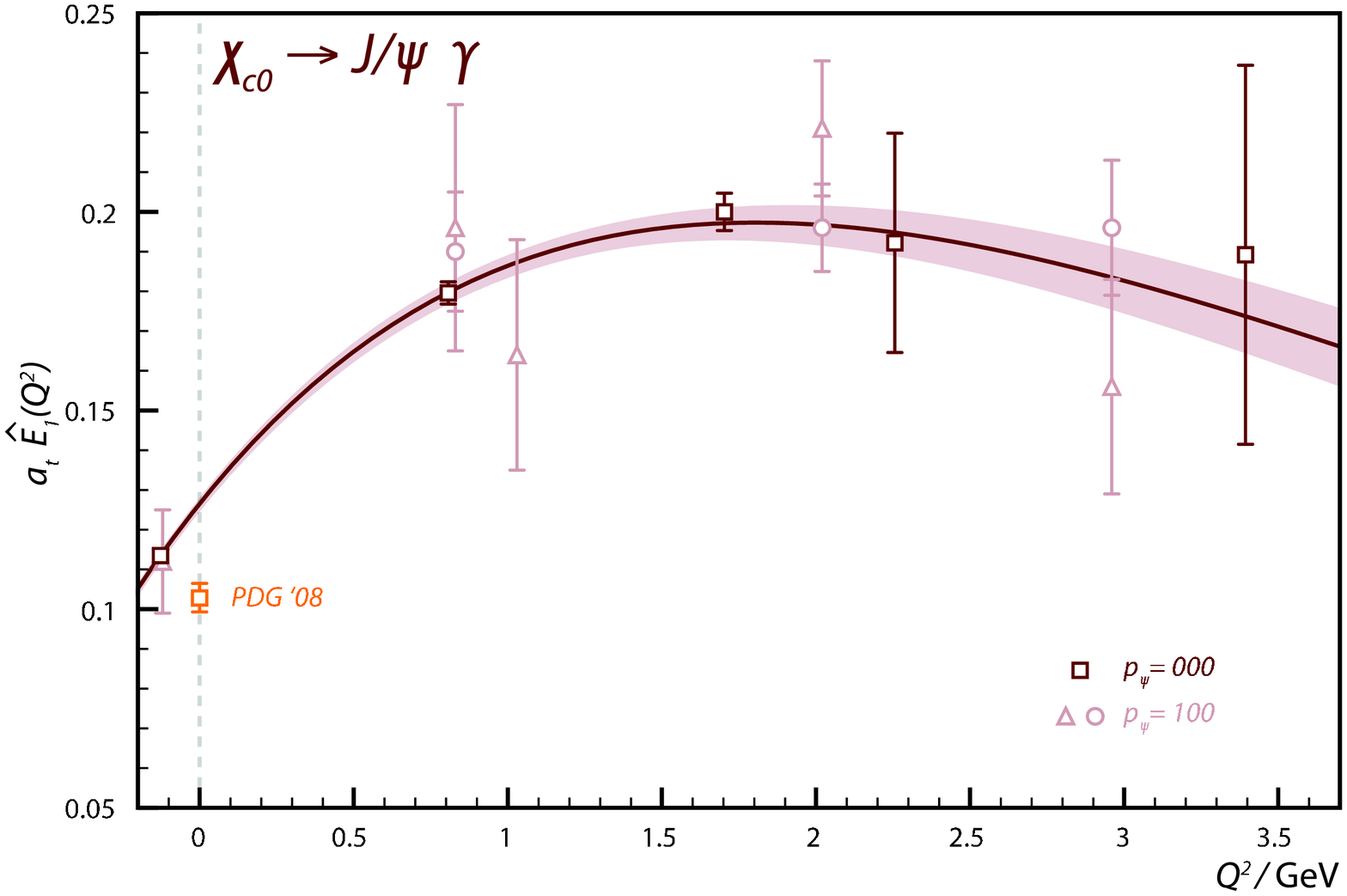}
 \includegraphics[width=8cm,bb=0 0 683 480]{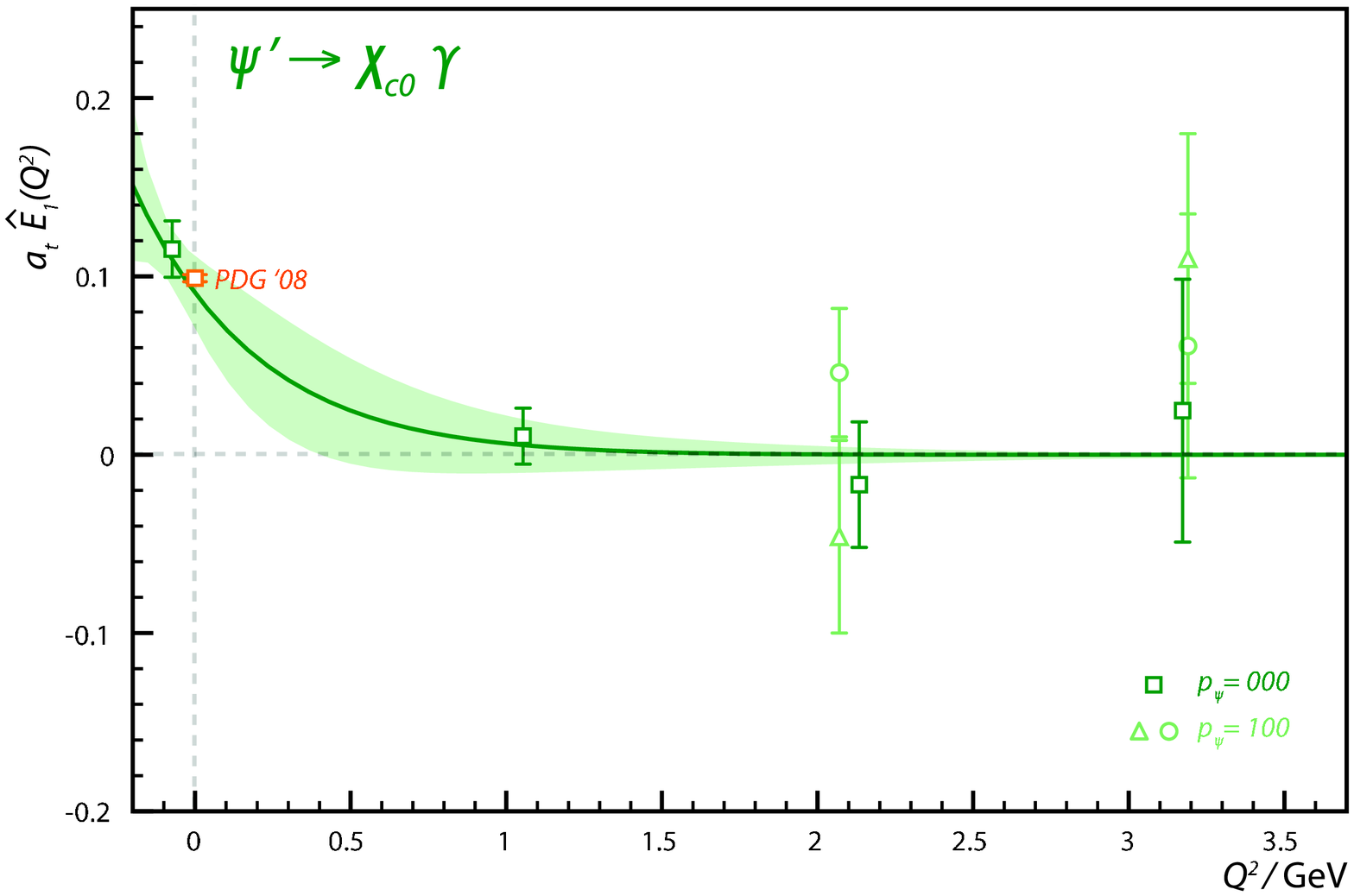}
\includegraphics[width=8cm,bb=0 0 680 480]{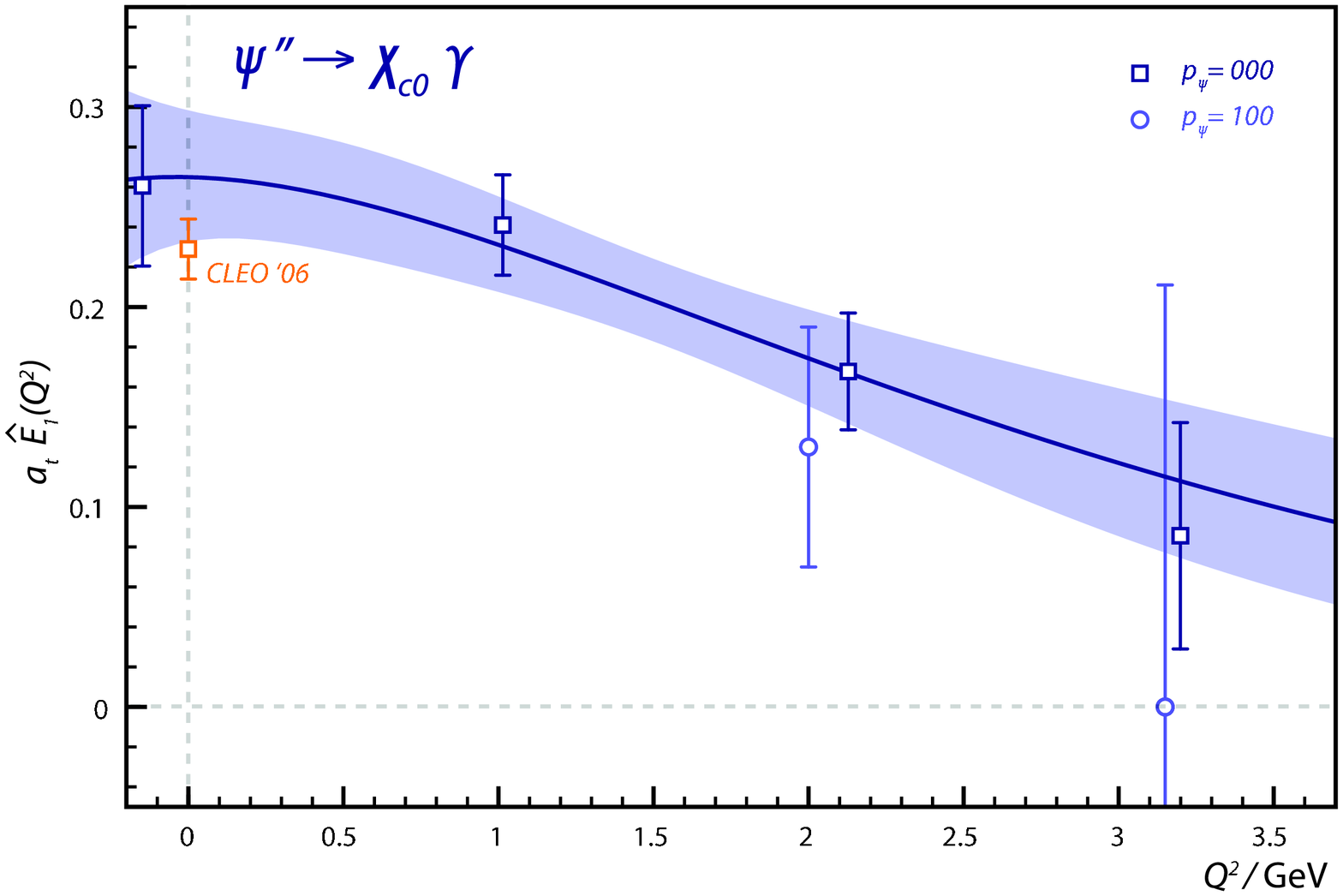}
\includegraphics[width=8cm,bb=0 0 680 480]{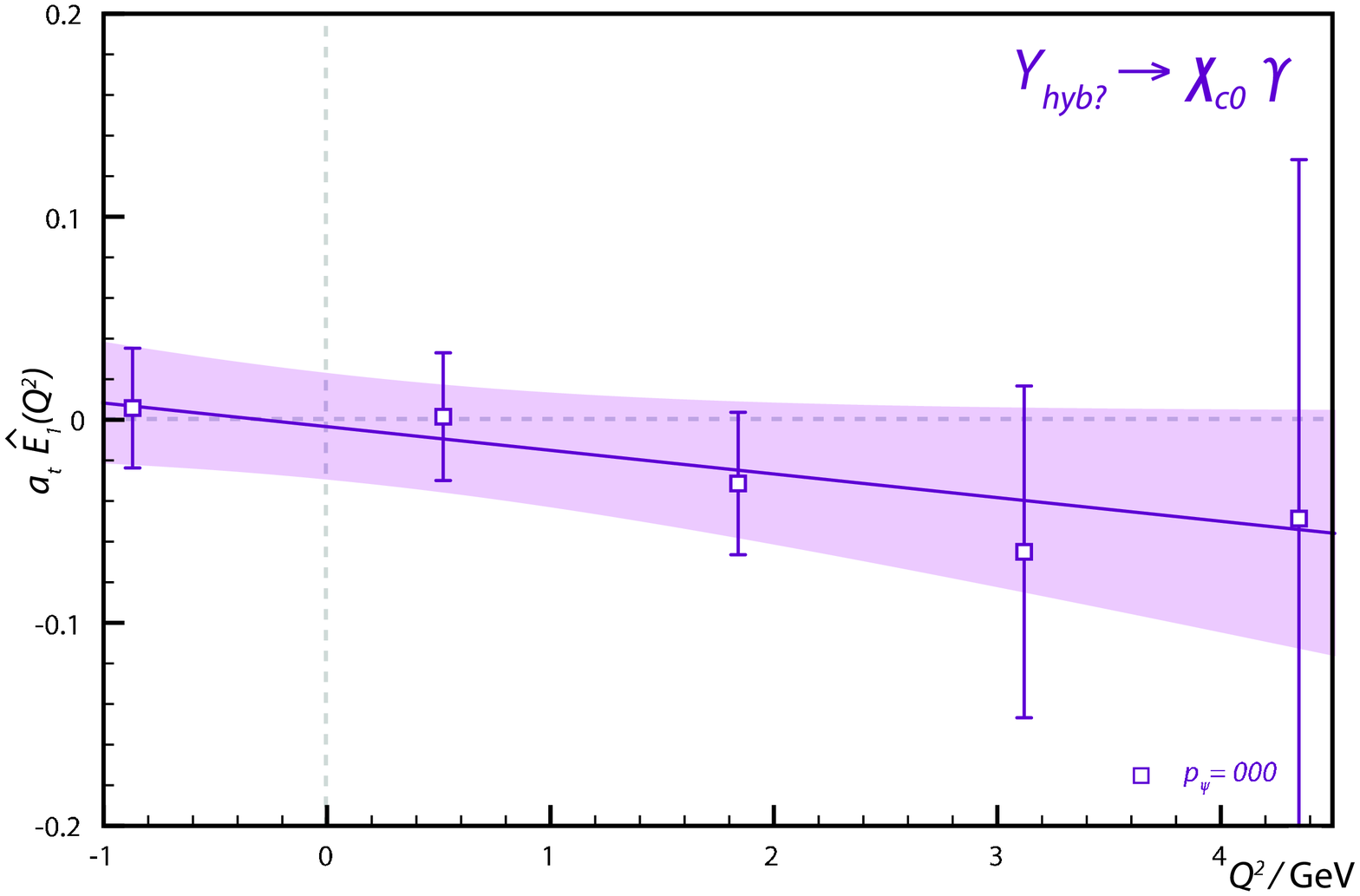}
\caption{\label{SVfig}Electric dipole transition form-factors $\chi_{c0} \leftrightarrow \psi$. Plotted is the form-factor in temporal lattice units against the photon virtuality in $\mathrm{GeV}^2$. Fits to the lattice $Q^2$ dependence as described in the text. Experimental points at $Q^2=0$ are extracted from experimental decay widths taken from \cite{Amsler:2008zzb, Briere:2006ff}.} 
\end{figure} 

In each case the $Q^2$ dependence of the lattice data has been fitted with the form,
\begin{equation}
 	E_1(Q^2) = E_1(0) \left( 1 + \lambda Q^2 \right) e^{-\frac{Q^2}{16\beta^2}},	\label{poly_times_exp}
\end{equation}
whose motivation is described in \cite{Dudek:2006ej}. The results of the fits are given in Table \ref{SVfit}. 
One point to make here is that in our lattice calculations, the photon only couples to the quark and not the antiquark and we do not explicitly include the electric charge of the quark.  Therefore we actually compute and present $\hat{F}_k(Q^2)$  which are related to the physical multipole amplitudes by $F_k(Q^2) = 2 \times \frac{2}{3}e \times \hat{F}_k(Q^2)$. Note also that the multipole amplitudes have mass dimension 1 and hence we plot them in temporal lattice units, where $a_t^{-1} = 6.05$ GeV is determined from the static potential. Partial decay widths follow by averaging over initial helicities and summing over final helicities which gives, for a general $A\to B \gamma$ decay:
$$
\Gamma(A \to B \gamma) = \frac{1}{2J_A + 1} \alpha \frac{16}{9} \frac{|\vec{q}|}{ m_A^2} \sum_k \left| \hat{F}_k(0)\right|^2
$$
Where we have experimental masses for states we use these to compute the phase space, otherwise we use the value extracted from the lattice calculation\cite{Dudek:2007wv, Dudek:2008sz}.

The ground state transition $\chi_{c0} \to J/\psi \gamma$ form-factor shows behavior rather similar to that found in \cite{Dudek:2006ej} which used the same lattices but Domain Wall fermions rather than Clover and which did not make use of multiple sink-operators and operator projection. We will consider this in a little more detail in Appendix B where we will consider the effect of ${\cal O}(ma)$ improvement of the local vector current. We will discuss the results in comparison with experiment and with quark potential models in section \ref{phenom}.

\begin{table}[t]
 \begin{tabular}{c c| c  c  | cc}
$\begin{matrix}\mathrm{sink} \\ \mathrm{level}\end{matrix}$ & $\begin{matrix}\mathrm{suggested} \\ \mathrm{transition}\end{matrix}$  & $a_t \hat{E}_1(0)$ & $\begin{matrix}\beta/\mathrm{MeV} \\ \lambda/\mathrm{GeV^{-2}} \end{matrix}$ & $\Gamma_{\mathrm{lat}}/\mathrm{keV}$ & $\Gamma_{\mathrm{expt}}/\mathrm{keV}$\\
\hline
0 & $\chi_{c0} \to J/\psi \gamma$ & $0.127(2)$ & $\begin{matrix} 409(12)\\ 1.14(5) \end{matrix}$  & $199(6)$  & $131(14)$  \\
1 & $\psi' \to \chi_{c0} \gamma$ & $0.092(19)$ & $\begin{matrix}164(55) \\ \mathrm{0[fixed]} \end{matrix} $ &   $26(11)$  & $30(2)$\\
3 & $\psi'' \to \chi_{c0} \gamma$ & $0.265(33)$ & $\begin{matrix}324(77) \\ 0.58(56) \end{matrix}$ & $265(66)$ & $199(26)$ \\
5 & $Y_{\mathrm{hyb.}} \to \chi_{c0} \gamma$ & $0.00(3)$ &$\begin{matrix}\mathrm{linear}\\ \mathrm{fit}\end{matrix}$  & $\lesssim 20$ & -
 \end{tabular} 
\caption{\label{SVfit}Results of fit to lattice data using equation \ref{poly_times_exp}. Partial decay width computed using fitted value of $E_1(0)$ and physical phase space (where known). All errors are just lattice statistical. Experimental partial decay widths from \cite{Amsler:2008zzb, Briere:2006ff}.}
\end{table} 

\subsection{Vector - Pseudoscalar transitions}
\begin{figure}[!]
\includegraphics[width=8cm,bb=0 0 680 480]{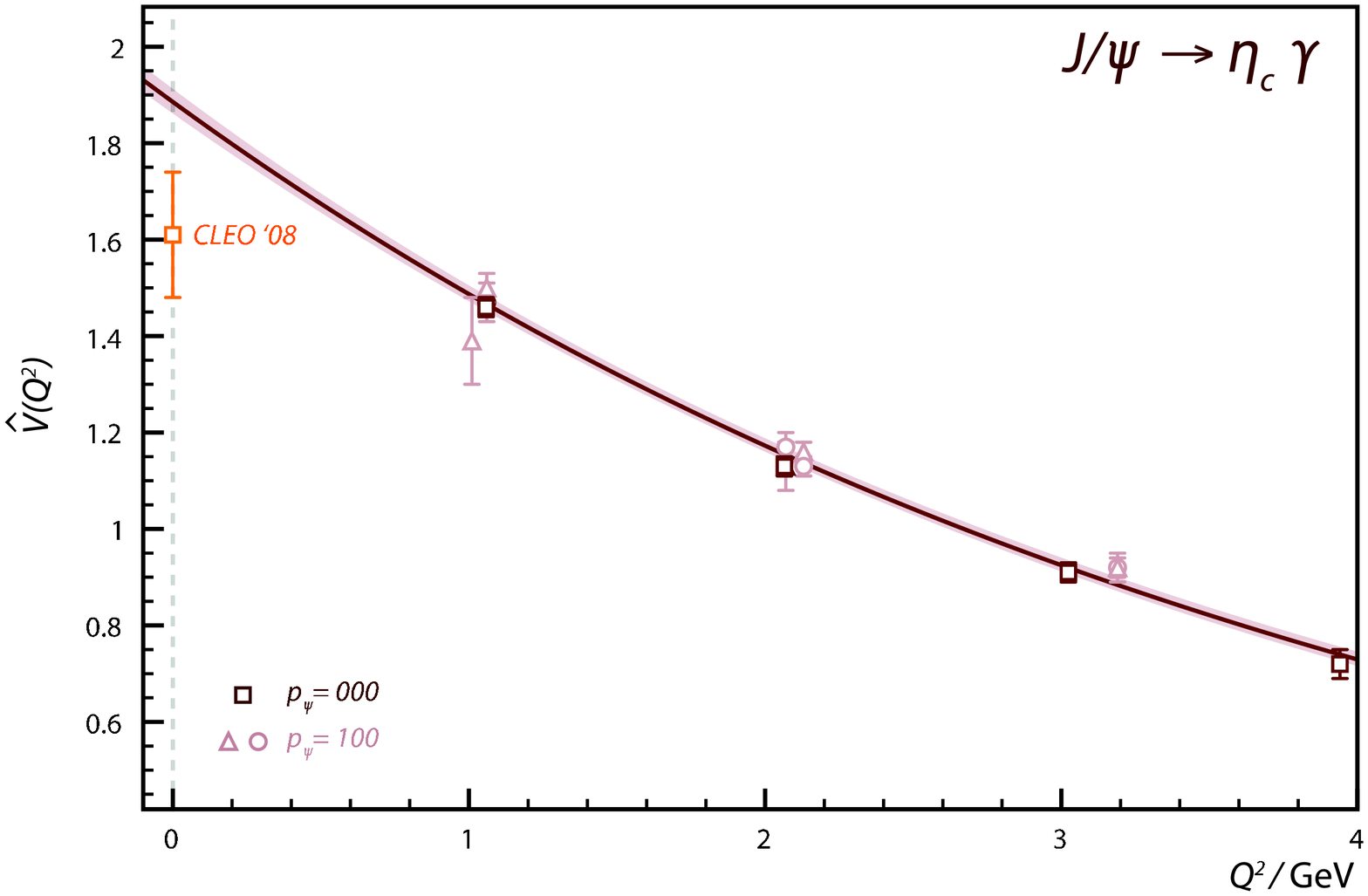}
\includegraphics[width=8cm,bb=0 0 680 480]{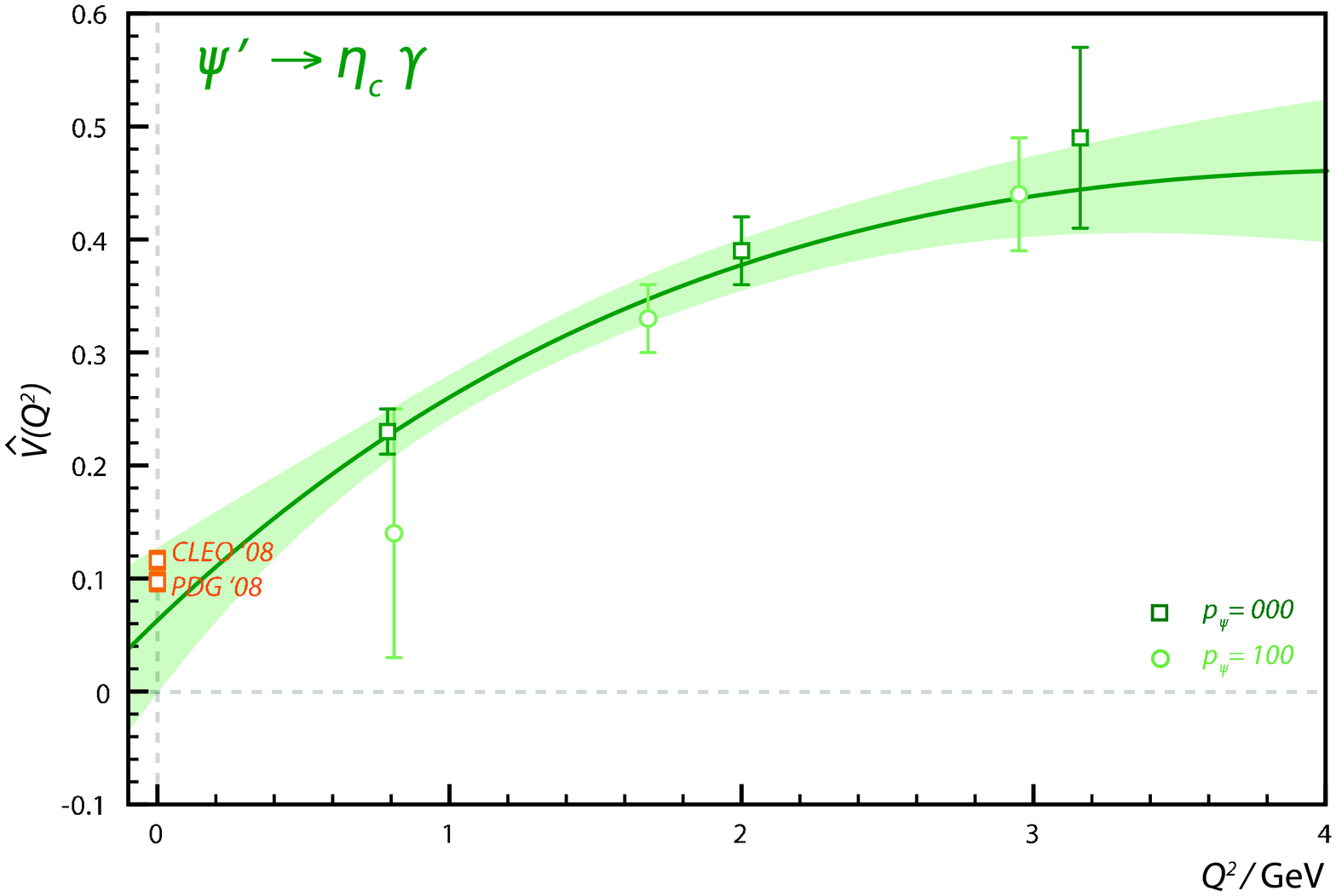}
\includegraphics[width=8cm,bb=0 0 680 480]{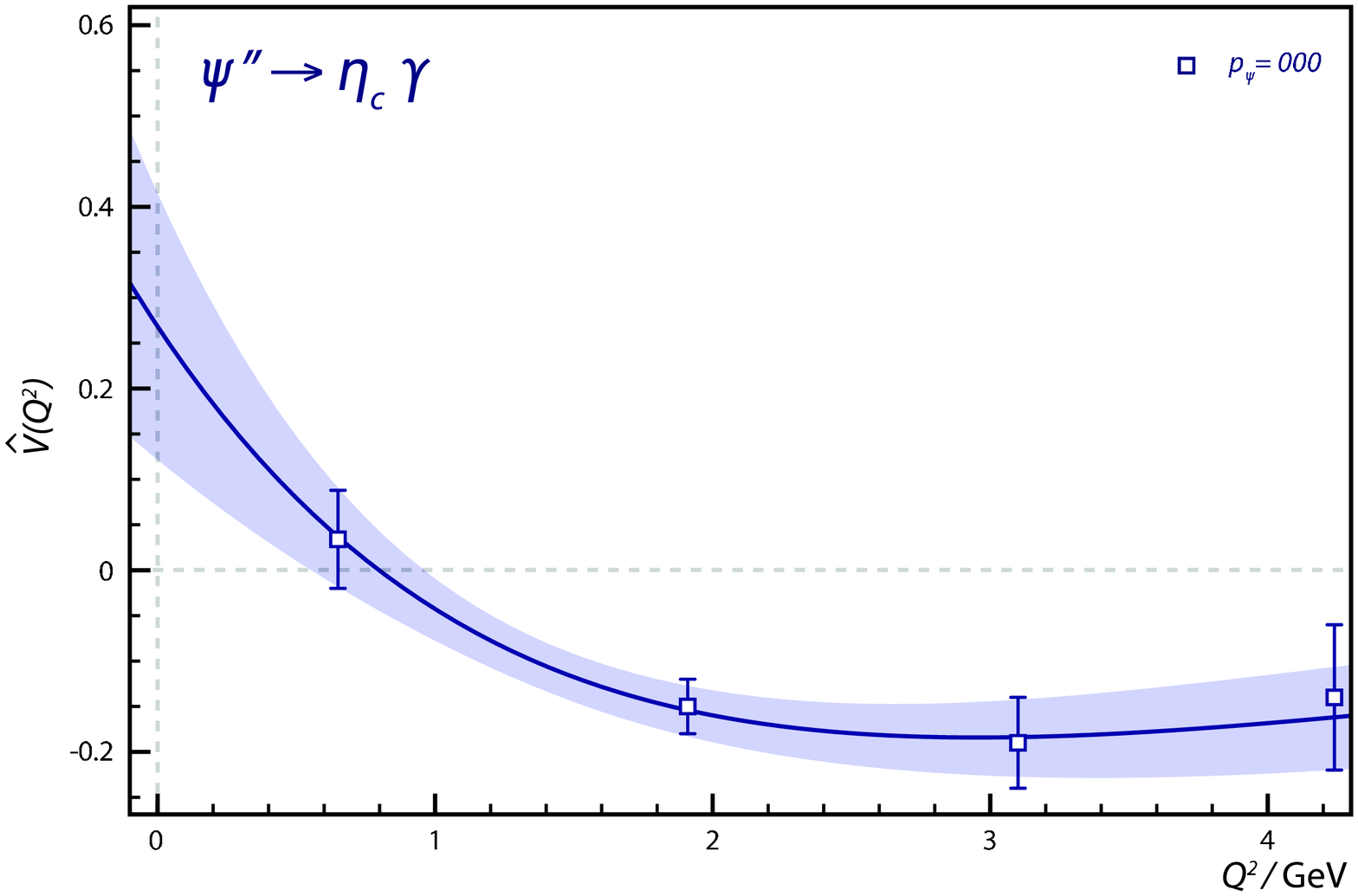}
\includegraphics[width=8cm,bb=0 0 680 480]{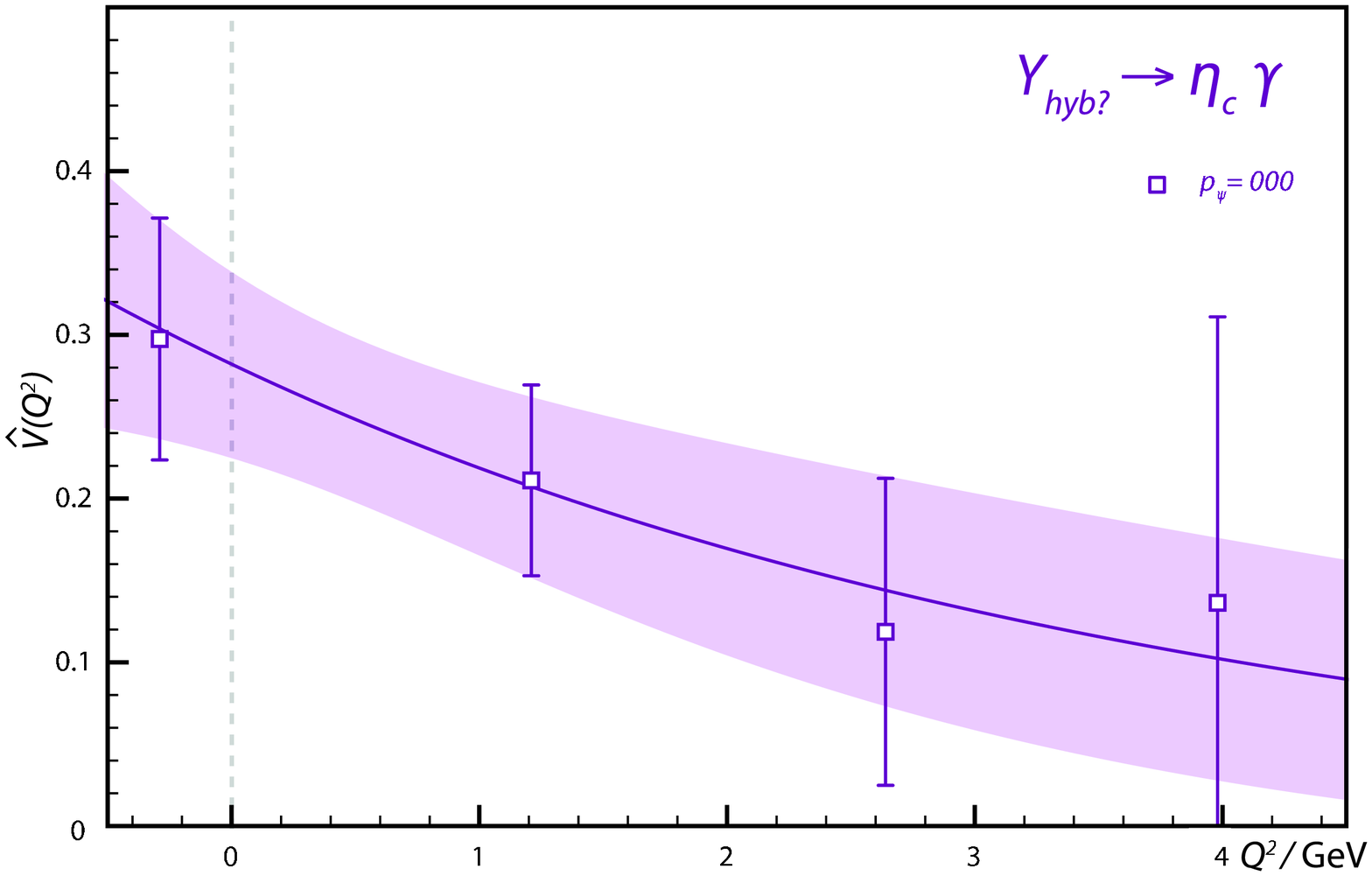}
\caption{\label{PVfig}Magnetic dipole transition form-factors $\psi \to \eta_c$. Plotted is the dimensionless form-factor against the photon virtuality in $\mathrm{GeV}^2$. Fits to the lattice $Q^2$ dependence as described in the text. Experimental points at $Q^2=0$ are extracted from experimental decay widths taken from \cite{Amsler:2008zzb, Mitchell:2008fb}. }
\end{figure}

Using the same $T_1^{--}$ operator set at the sink and the quark-smeared $\bar{\psi} \gamma^5 \psi$ operator at the source we obtained results for the single magnetic dipole form-factor in the vector-pseudoscalar transition. The decomposition used\footnote{note that this is not the conventionally normalized magnetic dipole amplitude} is
\begin{equation} 
 \label{eq:rhopi} 
 \langle P(\vec{p}_P) | j^\mu(0) | V(\vec{p}_V, \lambda)\rangle =  \frac{2 V(Q^2)}{m_P + m_V}  \epsilon^{\mu\alpha\beta\gamma} p_{P\alpha} p_{V\beta}  \epsilon_\gamma(\vec{p}_V, \lambda). \nonumber
\end{equation} 
Figure \ref{PVfig} shows the form-factors for the transition between the lightest four vector states (ignoring the suspected $3^{--}$ intruder and the noisy $\psi($`4040'$)$ state) and the $\eta_c$. The fit-form in eqn \ref{poly_times_exp} was again used, with the fit results presented in table \ref{PVtab}. We refer the reader to the paper \cite{Dudek:2006ej} for a discussion of the systematic error introduced into the phase space by an inaccurate lattice estimate of the hyperfine splitting.

\begin{table}
 \begin{tabular}{c c| c   c | cc}
$\begin{matrix}\mathrm{sink} \\ \mathrm{level}\end{matrix}$ & $\begin{matrix}\mathrm{suggested} \\ \mathrm{transition}\end{matrix}$  & $\hat{V}(0)$ & $\begin{matrix}\beta/\mathrm{MeV} \\ \lambda/\mathrm{GeV^{-2}} \end{matrix}$& $\Gamma_{\mathrm{lat}}$/keV & $\Gamma_{\mathrm{expt}}$/keV\\
\hline
0 & $J/\psi \to \eta_c \gamma$ & $1.89(3)$ & $\begin{matrix} 513(7) \\ \mathrm{0[fixed]}\end{matrix}$ & $2.51(8)$ & $1.85(29)$ \\
1 & $\psi' \to \eta_c \gamma$ & $0.062(64)$ & $\begin{matrix} 530(110) \\ 4(6)\end{matrix}$ & $0.4(8)$ & $\begin{matrix}0.95(16)\\ 1.37(20)\end{matrix}$ \\
3 & $\psi'' \to \eta_c \gamma$ & $0.27(15)$ & $\begin{matrix} 367(55) \\ -1.25(30)\end{matrix}$ & $10(11)$ & - \\
5 & $Y_{\mathrm{hyb.}} \to \eta_c \gamma$ & $0.28(6)$ & $\begin{matrix} 250(200) \\ \mathrm{0[fixed]}\end{matrix}$ & $42(18)$ & - 
 \end{tabular} 
\caption{\label{PVtab}Results of fit to lattice data using equation \ref{poly_times_exp}. Partial decay width computed using fitted value of $\hat{V}(0)$ and physical phase space (where known). All errors are purely lattice statistical. Experimental partial decay widths from \cite{Amsler:2008zzb, Mitchell:2008fb}.}
\end{table} 

\begin{figure}[h!]
 \includegraphics[width=9.5cm,bb=0 0 661 439]{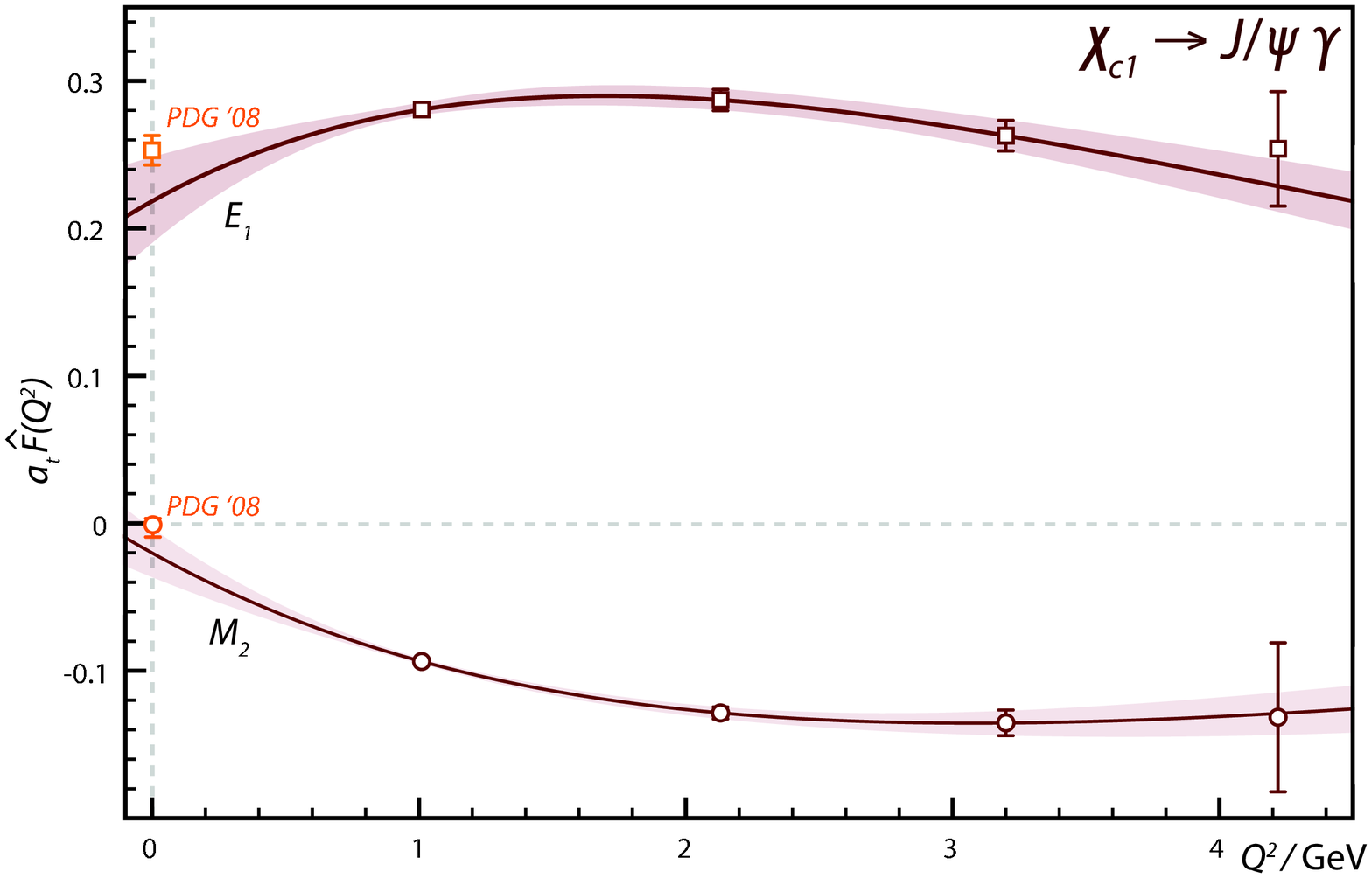}
\includegraphics[width=9.5cm,bb=0 0 661 439]{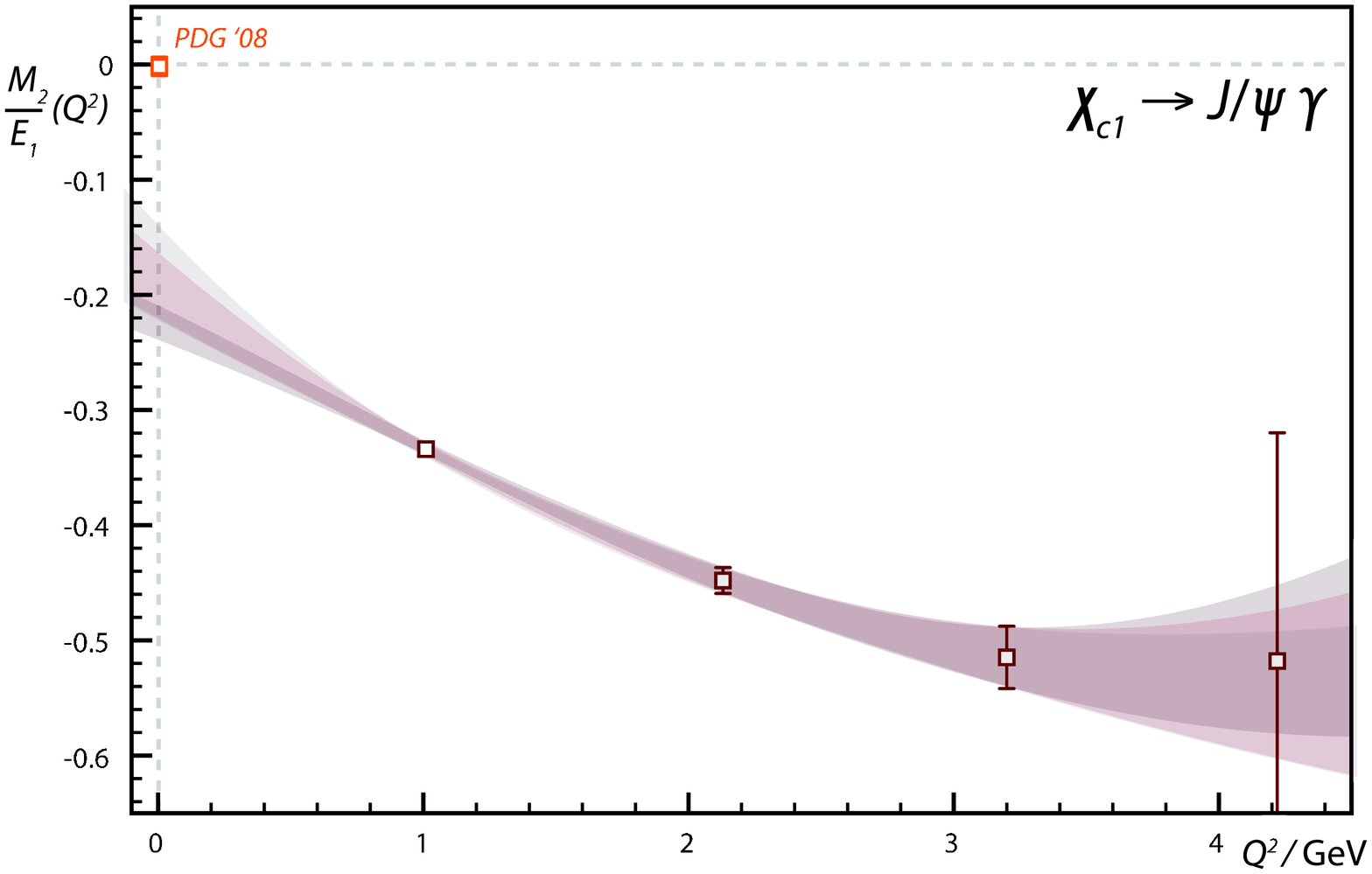}	
\caption{\label{axialvector}(a) Electric dipole and magnetic quadrupole form-factors for the transition $\chi_{c1} \to J/\psi \gamma$. $Q^2$ dependence fitted with equation \ref{poly_times_exp} and extrapolated to the physical photon point $Q^2=0$ for comparison with experimental data from \cite{Amsler:2008zzb}. Relative sign of $E_1$ to $M_2$ is relevant. (b) Ratio of magnetic quadrupole to electric dipole form-factors. Colored bands represent fits with various fit functions.}
\end{figure} 

\begin{figure}[h!]
 \includegraphics[width=9.5cm,bb=0 0 661 439]{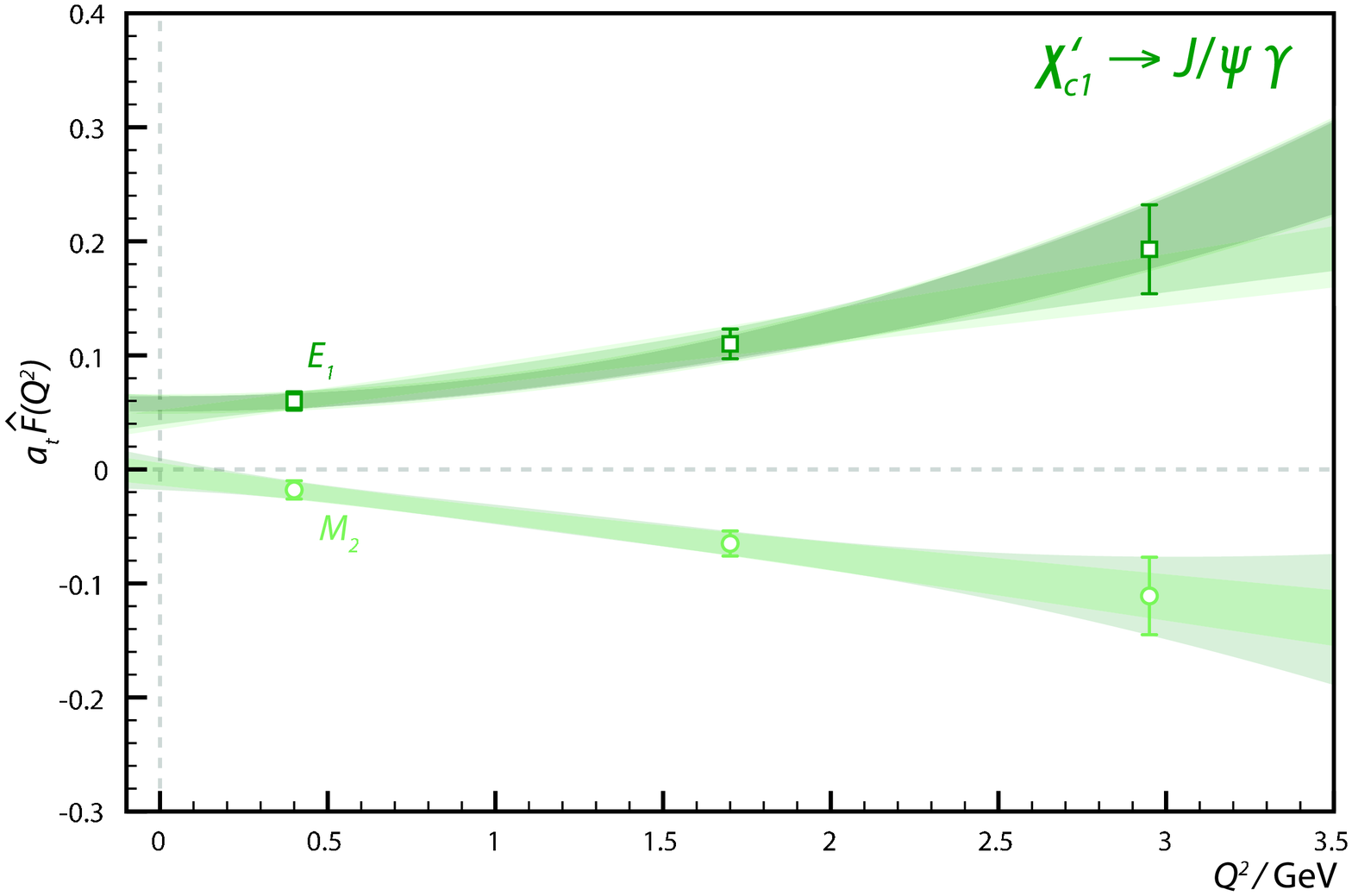}
\caption{\label{axialvector_excited}Electric dipole and magnetic quadrupole form-factors for the transition $\chi_{c1}' \to J/\psi \gamma$. $Q^2$ dependence fitted with various forms shown by the colored bands. Relative sign of $E_1$ to $M_2$ is relevant. }
\end{figure}

\subsection{Axial - Vector transitions}
In transitions between axial ($1^{++}$) and vector ($1^{--}$) states there are two transverse form-factors, electric dipole ($E_1$) and magnetic quadrupole ($M_2$), and one longitudinal form-factor ($C_1$). The matrix element decomposition takes the form

\begin{align}  
&\hspace{-1cm}\langle A(\vec{p}_A, \lambda_A) | j^\mu(0) | V(\vec{p}_V, \lambda_V)\rangle = \tfrac{i}{4\sqrt{2} \Omega(Q^2)} \epsilon^{\mu\nu\rho\sigma} (p_A  - p_V)_\sigma  \nonumber \\  
&\hspace{-1cm} \quad\times \Bigg[ E_1(Q^2) (p_A+p_V)_\rho \Big( 2m_A  [\epsilon^*(\vec{p_A}, \lambda_A) \cdot p_V] \epsilon_\nu(\vec{p_V}, \lambda_V)  + 2m_V [\epsilon(\vec{p}_V, \lambda_V) \cdot p_A]  \epsilon^*_\nu(\vec{p}_A, \lambda_A)  \Big) \nonumber\\
&\hspace{-1cm} \quad\quad\quad+ M_2(Q^2) (p_A+p_V)_\rho \Big( 2m_A  [\epsilon^*(\vec{p_A}, \lambda_A) \cdot p_V] \epsilon_\nu(\vec{p_V}, \lambda_V)  - 2m_V [\epsilon(\vec{p}_V, \lambda_V) \cdot p_A]  \epsilon^*_\nu(\vec{p}_A, \lambda_A)  \Big) \nonumber\\
&\hspace{-1cm} \quad\quad\quad+ \frac{C_1(Q^2)}{\sqrt{q^2}} \Big( - 4\Omega(Q^2) \epsilon^*_\nu(\vec{p}_A, \lambda_A) \epsilon_\rho( \vec{p}_V,\lambda_V)   \nonumber \\
&\hspace{-1cm}\quad\quad\quad\quad + (p_A+p_V)_\rho \Big[ (m_A^2-m_V^2 + q^2) [\epsilon^*(\vec{p}_A, \lambda_A) \cdot p_V]\; \epsilon_\nu(\vec{p}_V, \lambda_V) + (m_A^2 - m_V^2 -q^2) [\epsilon(\vec{p}_V, \lambda_V).p_A] \; \epsilon^*_\nu(\vec{p_A}, \lambda_A) \Big]         \Big)\Bigg]. \nonumber
\end{align}

In this case\footnote{We could have used $\bar{\psi} \gamma^5 \gamma^j \psi$ at the source to produce the axial meson, this operator at finite momentum also has overlap with the pseudoscalar state which severely limits its usefulness.} we computed with the vector at the source, produced using quark-smeared $\bar{\psi} \gamma^j \psi$, and a set of ten axial (actually $T_1^{++}$) operators at the sink\footnote{$\gamma_i \gamma_5$, $\rho\times \nabla_{T1}$, $\rho_{(2)}\times \nabla_{T1}$, $a_1\times \mathbb{D}_{T1}$, $b_1\times\mathbb{B}_{T1}$ in both quark-smeared and unsmeared versions.}.  In figure \ref{axialvector}(a) we present the two transverse form-factors for the ground state axial meson and in figure \ref{axialvector}(b) their ratio. The amplitudes were each fitted with a form like eqn. \ref{poly_times_exp} which unfortunately is not as well constrained as in the scalar case owing to the kinematic factors preventing a slightly time-like $Q^2$ point corresponding to $\vec{p}_i = \vec{p_f} = (000)$. The results of the fits are $a_t \hat{E}_1(0) = 0.23(3)$, $\beta_{E1} = 440(40)\,\mathrm{MeV}$, $\lambda_{E1} = 0.71(30)\, \mathrm{GeV}^{-2}$ and $a_t \hat{M}_2(0) = -0.020(17)$, $\beta_{M2} = 450(50)\,\mathrm{MeV}$, $\lambda_{M2} = 5(6) \, \mathrm{GeV}^{-2}$. This corresponds to a partial decay width of $\Gamma(\chi_{c1} \to J/\psi \gamma) =270(70) \,\mathrm{keV}$ which is in reasonable agreement with the PDG's average of  $320(25)\,\mathrm{keV}$.

The ratio $\frac{M_2}{E_1}(Q^2)$, shown in figure \ref{axialvector}(b), was fitted with various functional forms shown by the shaded bands yielding $\frac{M_2}{E_1}(0) =
-0.20(6)$ where the error includes a crude estimate of the systematic error
due to the uncertainty in fitting form. The ratio of the extrapolated values from the separate fits to $E_1$, $M_2$ gives $\frac{M_2}{E_1}(0) = \frac{-0.020(17)}{0.23(3)}  = -0.09(7) $. Clearly without data points at
smaller $Q^2$ or some certainty about the expected $Q^2$ dependence, we cannot constrain this any further and hence cannot make a particularly meaningful comparison with the PDG average $\frac{M_2(0)}{\sqrt{E_1(0)^2 + M_2(0)^2}} = - 0.002 \substack{+0.008 \\ -0.017}$.

Form-factors for the transition from the first excited axial state, $\chi_{c1}'$ down to the $J/\psi$ are shown in figure \ref{axialvector_excited} where multiple fit forms were used, all returning a $\chi^2/N_\mathrm{dof}$ close to one. The estimates for the physical photon point thus obtained are $a_t \hat{E}_1(0) = 0.050(15)$ and $a_t \hat{M}_2(0) = -0.004(14)$ where again we include a crude systematic error estimate for the fit-form variation. The $E_1$ transition corresponds, for a $\chi_{c1}'$ at $4.1$ GeV, to a partial decay width $\Gamma(\chi_{c1}' \to J/\psi \gamma) = 21(12)\,\mathrm{keV}$.

\subsection{Tensor - Vector transition}
The transition between the lightest $2^{++}$ state and the lightest vector state was not considered in \cite{Dudek:2006ej} as that study used only local fermion bilinears ($\bar{\psi}(x) \Gamma \psi(x)$) to produce states - a spin-2 particle cannot be produced by any such operator. Here we use a set of six operators projected into $T_2^{++}$ and $E^{++}$ irreps at the sink\footnote{$\rho\times\nabla$, $\rho_{(2)}\times\nabla$, $b_1\times \mathbb{B}$ in both quark smeared and unsmeared versions.}. The multipole decomposition for this transition takes the following form, 
\begin{widetext}
\begin{align}
\langle V&(\vec{p}_V, \lambda_V) | j^{\mu}(0) | T(\vec{p}_T, \lambda_T) \rangle = \nonumber \\
& E_1(Q^2) \sqrt{\frac{3}{5}} \Bigg[ -A^{\mu} + \frac{m_T}{\Omega}\left(\tilde{\omega}-m_V\right)B^{\mu}
+ \frac{m_T}{\Omega}\Big( \tilde{\omega} D_T^{\mu} - m_T D_V^{\mu} \Big)
+ \frac{m_T^2}{\Omega^2}(\tilde{\omega}-m_V)\Big( -\tilde{\omega} F_T^{\mu} + m_T F_V^{\mu} \Big)  \Bigg]\nonumber \\
& + M_2(Q^2) \sqrt{\frac{1}{3}} \Bigg[ A^{\mu} - \frac{m_T}{\Omega}\left(\tilde{\omega}+m_V\right)B^{\mu}
- \frac{2 m_T^2}{\Omega} C^{\mu} + \frac{m_T}{\Omega}\Big( -\tilde{\omega} D_T^{\mu} + m_T D_V^{\mu} \Big)\nonumber \\
&\quad\quad\quad\quad\quad\quad + \frac{m_T^2}{\Omega^2}\Big( \left(\tilde{\omega}^2 + \tilde{\omega} m_V - 2m_V^2\right) F_T^{\mu} + m_T\left(\tilde{\omega}-m_V\right)F_V^{\mu} \Big)  \Bigg]\nonumber \\
& + E_3(Q^2) \sqrt{\frac{1}{15}} \Bigg[ -A^{\mu} + \frac{m_T}{\Omega}\left(\tilde{\omega}+4m_V\right)B^{\mu}
- \frac{5 m_T^2}{2\Omega} C^{\mu} + \frac{m_T}{\Omega}\Big( \tilde{\omega} D_T^{\mu} - m_T D_V^{\mu} \Big) \nonumber\\
&\quad\quad\quad\quad\quad\quad + \frac{m_T^2}{\Omega^2}\Big( -\left(\tilde{\omega}^2 + 4 \tilde{\omega} m_V + \tfrac{5}{2}m_V^2\right) F_T^{\mu} + m_T\left(\tfrac{7}{2}\tilde{\omega}+4m_V\right)F_V^{\mu} \Big)  \Bigg] \nonumber\\
& + C_1(Q^2) \sqrt{\frac{3}{5}}\frac{m_T}{\Omega \sqrt{q^2}} \Bigg[
 \left(m_V^2-\tilde{\omega} m_T \right) D_T^{\mu} + \left(m_T^2 - \tilde{\omega} m_T \right) D_V^{\mu} \nonumber\\
& \quad\quad\quad\quad\quad\quad\quad -\frac{m_T}{\Omega}\big(\tilde{\omega}-m_V\big) \Big(\left(m_V^2-\tilde{\omega} m_T \right) F_T^{\mu} + \left(m_T^2-\tilde{\omega}  m_T \right) F_V^{\mu} \Big) \Bigg]\nonumber \\
& + C_3(Q^2) \sqrt{\frac{2}{5}}\frac{m_T}{\Omega \sqrt{q^2}} \Bigg[
 \left(m_V^2-\tilde{\omega} m_T \right) D_T^{\mu} + \left(m_T^2-\tilde{\omega} m_T \right) D_V^{\mu}\nonumber \\
& \quad\quad\quad\quad\quad\quad\quad - \frac{m_T}{\Omega}\big(\tilde{\omega} +\tfrac{3}{2}m_V\big) \Big(\left(m_V^2-\tilde{\omega} m_T \right) F_T^{\mu} + \left(m_T^2-\tilde{\omega}  m_T \right) F_V^{\mu} \Big) \Bigg].
\label{TV}
\end{align}
Here $\Omega \equiv (p_T\cdot p_V)^2 - m_T^2 m_V^2$ and $\tilde{\omega} \equiv \tfrac{p_V \cdot p_T}{ m_T}$ and
\begin{multline}
A^{\mu} \equiv \epsilon^{\mu \nu}(\vec{p}_T, \lambda_T) \epsilon^*_{\nu}(\vec{p}_V, \lambda_V);\; 
B^{\mu} \equiv \epsilon^{\mu \nu}(\vec{p}_T, \lambda_T) p^V_{\nu} (\epsilon^*(\vec{p}_V, \lambda_V) \cdot p_T);\;
C^{\mu} \equiv \epsilon^{*\mu}(\vec{p}_V, \lambda_V) (\epsilon^{\alpha \beta}(\vec{p}_T, \lambda_T) p^V_{\alpha} p^V_{\beta}); \nonumber\\
D_T^{\mu} \equiv p_T^{\mu} (\epsilon^{\alpha \beta}(\vec{p}_T, \lambda_T) \epsilon^*_{\alpha}(\vec{p}_V, \lambda_V) p^V_{\beta});\;
D_V^{\mu} \equiv p_V^{\mu} (\epsilon^{\alpha \beta}(\vec{p}_T, \lambda_T) \epsilon^*_{\alpha}(\vec{p}_V, \lambda_V) p^V_{\beta});\nonumber \\
F_T^{\mu} \equiv p_T^{\mu} (\epsilon^{\alpha \beta}(\vec{p}_T, \lambda_T) p^V_{\alpha} p^V_{\beta}) (\epsilon^*(\vec{p}_V, \lambda_V) \cdot p_T);\;
F_V^{\mu} \equiv p_V^{\mu} (\epsilon^{\alpha \beta}(\vec{p}_T, \lambda_T) p^V_{\alpha} p^V_{\beta}) (\epsilon^*(\vec{p}_V, \lambda_V) \cdot p_T)  \nonumber
\end{multline}
\end{widetext}

In order to have a constrained linear system for extraction of the five form-factors we are required to consider all five spin-2 helicities, which with respect to the lattice cubic symmetry are distributed in $T_2(3)$ and $E(2)$ irreps. Strictly speaking these are independent irreps on the lattice and we should be careful about combining them, but any lattice symmetry breaking must vanish as $a \to 0$ and in \cite{Dudek:2007wv} we found that there were strong signals that the continuum rotational symmetry was restored to a good approximation already at this value of $a$. For example, for the ground states in the $T_2, E$ channels we have a high degree of mass degeneracy, and the values of the overlaps ($Z_{T_2}/Z_E$) are compatible at the $1\%$ level. Details of this analysis are in \cite{Dudek:2007wv}.

\begin{figure}[t!]
 \includegraphics[width=9.5cm,bb=0 0 661 439]{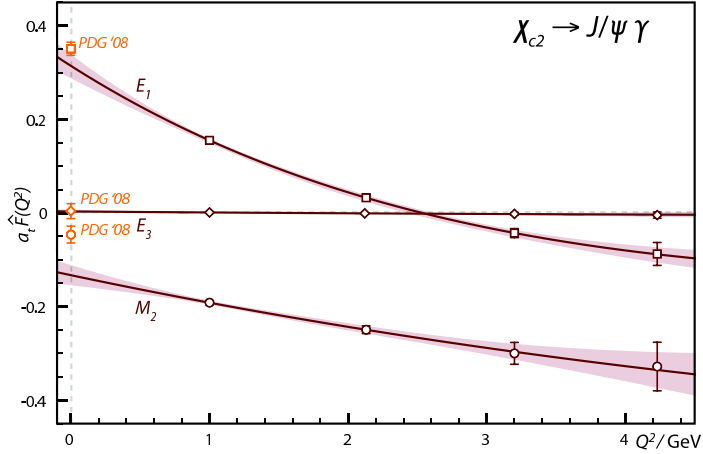}
\vspace{-8mm}
\caption{\label{tensorvector}Electric dipole, magnetic quadrupole and electric octopole form-factors for the transition $\chi_{c2} \to J/\psi \gamma$.  $Q^2$ dependencies fitted with equation \ref{poly_times_exp} and extrapolated to the physical photon point $Q^2=0$ for comparison with experimental data from \cite{Amsler:2008zzb}. Relative signs are relevant. }
\end{figure}

The extracted form-factors and fits to the $Q^2$ dependence are shown in figure \ref{tensorvector} where the expected hierarchy $|E_1(0)| > |M_2(0)| \gg   |E_3(0)|$, to be discussed in section \ref{phenom}, is observed. The description of the lattice three-point correlators by the continuum decomposition, eqn \ref{TV} (which does not take any account of $T_2/E$ discretization differences) is excellent, with typically $\chi^2/N_{\mathrm{dof}} \sim 1$. As a simple systematic test of the degree to which $T_2/E$ discretization differences could affect the determination of the form-factors we tried deliberately introducing such a difference by arbitrarily multiplying all $E$ correlators by a factor. Even with a $10\%$ increase in the magnitude of all $E$ correlators, the form-factor values changed by less than the statistical error, while the $\chi^2/N_{\mathrm{dof}}$ increased to around $6$. If we where to see such large $\chi^2$ in our raw data (which we never do) we would not trust the results - we conclude that our form-factor extractions are not being strongly affected by $T_2/E$ discretization differences.

The amplitudes at $Q^2=0$ correspond to a partial decay width of $\Gamma(\chi_{c2} \to J/\psi \gamma) = 380(50)$ keV to be compared with the PDG average of $406(31)$ keV. The electric dipole fit parameters are $\beta = 550(80)$ MeV and $\lambda = -0.39(1) \,\mathrm{GeV}^{-2}$. The fits yield values for the ratios of multipole amplitudes of $\frac{M_2(0)}{\sqrt{E_1(0)^2 + M_2(0)^2 + E_3(0)^2}} = -0.39(7)$ and  $\frac{E_3(0)}{\sqrt{E_1(0)^2 + M_2(0)^2 + E_3(0)^2}} = 0.010(11)$ which in the case of the first ratio is considerably larger than the PDG average of $-0.13(5)$ but which does appear to be of the correct sign.

\begin{figure}[t!]
 \includegraphics[width=9.5cm,bb=0 0 661 439]{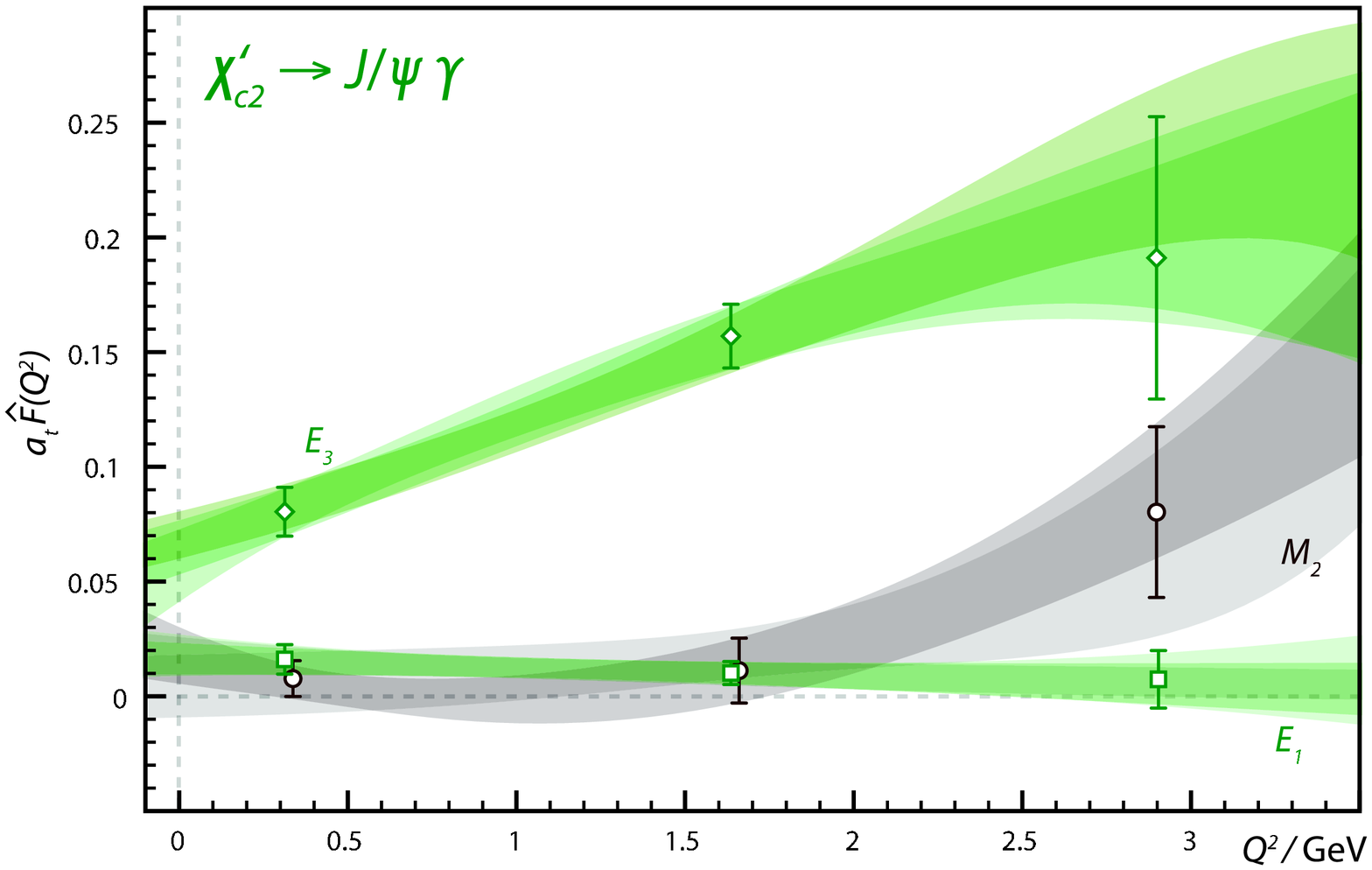}
 \includegraphics[width=9.5cm,bb=0 0 661 439]{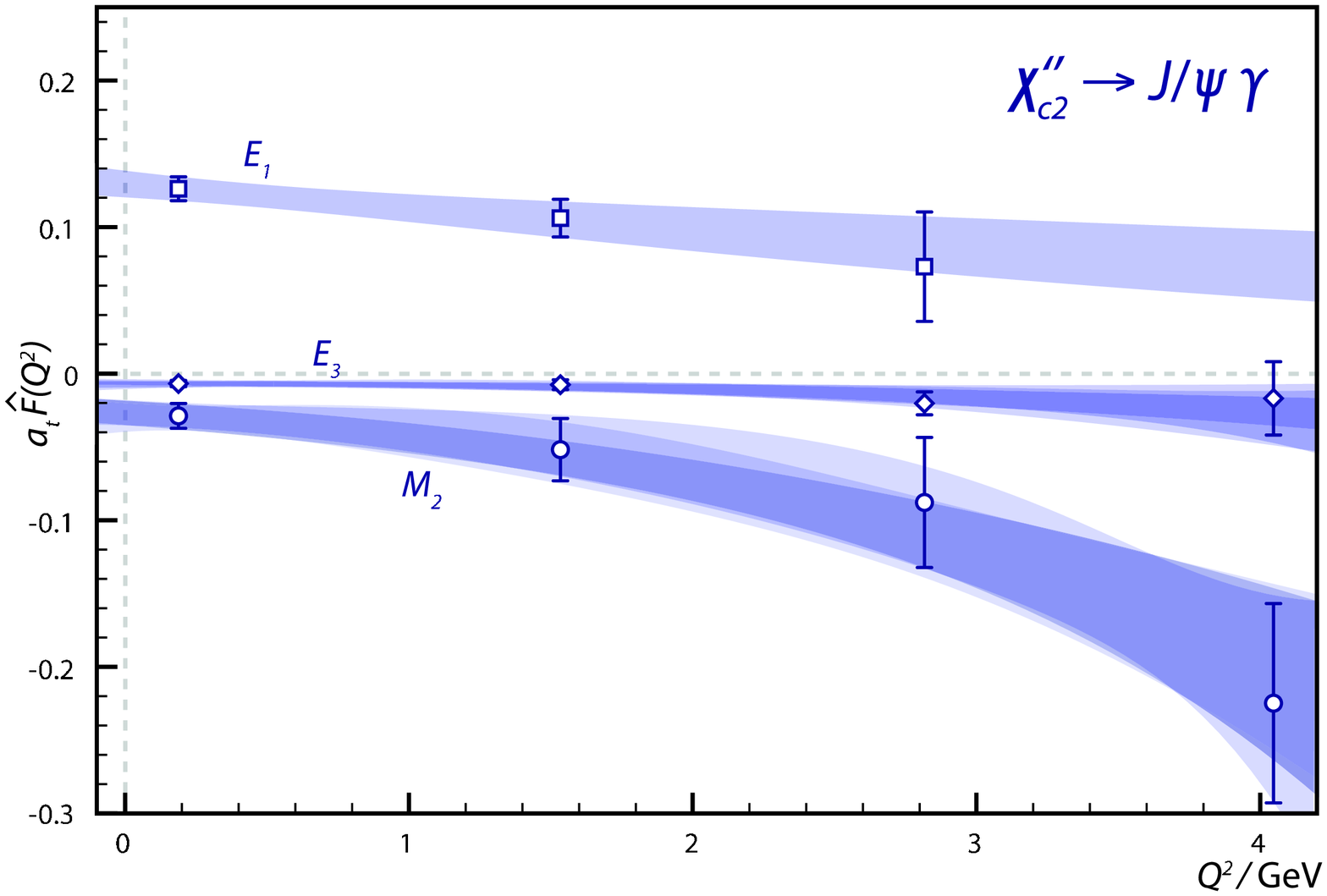}
\vspace{-8mm}
\caption{\label{tensorvector_exc}(a) Electric dipole, magnetic quadrupole and electric octopole form-factors for the transition $\chi_{c2}' \to J/\psi \gamma$.  $Q^2$ dependencies fitted with various fit-forms shown by the shaded bands. Relative signs are relevant. (b) Same for $\chi_{c2}'' \to J/\psi \gamma$.}
\end{figure} 

We also extracted the $\chi_{c2}' \to J/\psi \gamma$ transition form-factors\footnote{We found a mass of $4115(28)$ MeV for the $\chi_{c2}'$} as presented in figure \ref{tensorvector_exc}(a). The large value of $E_3$ might be surprising when compared to the ground state result - a number of simple systematic tests were performed to investigate if this might come about through various lattice effects with the result that we were unable to change the values outside statistical error bars by any reasonable adjustment. The amplitudes at $Q^2=0$ correspond to a partial decay width of $20(13)$ keV. In figure \ref{tensorvector_exc}(b) we present the transition form-factors for the next excited tensor state\footnote{We found a mass of $4165(30)$ MeV for the $\chi_{c2}''$}, $\chi_{c2}'' \to J/\psi \gamma$ where the hierarchy $|E_1(0)| \gg |M_2(0)|,|E_3(0)|$ appears to be restored and where we predict a partial decay width of $88(13)$ keV. In section \ref{phenom} we will propose a simple explanation for all these observations in the framework of a non-relativistic quark model.

\subsection{Exotic transitions}

Our principal focus here is the exotic $1^{-+}$ state, $\eta_{c1}$, here\footnote{But note our comments regarding the box-size as a cause of systematic error} found to be at $4300(50) $ MeV. On a quenched lattice, provided we can eliminate the possibility that the lightest state  in $T_1^{-+}$  is part of a non-exotic $4^{-+}$ (see \cite{Dudek:2008sz} for support of this elimination), we can be fairly certain that this state is a hybrid, having an excited gluonic field in addition to a charm-anticharm quark pair. This is in contrast to having a higher quark number Fock state which, since we lack light-quarks altogether in this calculation, could only arise for states having mass near $4m_c$\footnote{and in a quenched calculation they would arise in a unitarity-violating way \cite{Bardeen:2001jm}}.  A charge-conjugation allowed decay of this meson would be $\eta_{c1} \to J/\psi \gamma$ having transverse magnetic dipole and electric quadrupole multipole contributions. The decomposition of the vector current matrix element between two non-identical vector mesons takes the following form:
\begin{widetext}
\begin{align}
	\langle V'(\vec{p}', \lambda') |& j^\mu | V(\vec{p}, \lambda) \rangle = \nonumber  \\ - \left( \frac{m}{\sqrt{2\Omega}} \right) \Bigg[ & M_1(Q^2) \Bigg( \epsilon^{\mu*}(\vec{p}',\lambda')   \, \left( \epsilon(\vec{p},\lambda)\cdot p'\right) + \frac{m'}{m} \epsilon^\mu(\vec{p},\lambda) \, \left( \epsilon^*(\vec{p}',\lambda')\cdot p \right)  \nonumber \\ 
	&\quad\quad\quad\quad\quad +  \left( \epsilon(\vec{p},\lambda)\cdot p'\right) \, \left( \epsilon^*(\vec{p}',\lambda')\cdot p \right) \left(  \frac{q^\mu}{q^2}\left( \frac{m'}{m} - 1 \right)   + \frac{q^2 + \omega(m'-m)}{2 \Omega } \Pi^\mu    \right)        \Bigg) \nonumber  \\
	&+ E_2(Q^2) \Bigg(  \epsilon^{\mu*}(\vec{p}',\lambda')   \, \left( \epsilon(\vec{p},\lambda)\cdot p'\right) - \frac{m'}{m} \epsilon^\mu(\vec{p},\lambda) \, \left( \epsilon^*(\vec{p}',\lambda')\cdot p \right)  \nonumber  \\ 
	&\quad\quad\quad\quad\quad+  \left( \epsilon(\vec{p},\lambda)\cdot p'\right) \, \left( \epsilon^*(\vec{p}',\lambda')\cdot p \right) \left( -  \frac{q^\mu}{q^2}\left( \frac{m'}{m} + 1 \right)   + \frac{q^2 - \omega(m'+m)}{2 \Omega } \Pi^\mu    \right)        \Bigg) \nonumber  \\   	
	&+ \frac{C_0(Q^2)}{\sqrt{q^2}} \Bigg( \frac{q^2}{\sqrt{6} \, m} \Bigg) \Pi^\mu \Bigg( \left( \epsilon^*(\vec{p}',\lambda')\cdot \epsilon(\vec{p},\lambda) \right) + \frac{m(\omega + m' -m)}{\Omega}  \left( \epsilon(\vec{p},\lambda)\cdot p'\right) \, \left( \epsilon^*(\vec{p}',\lambda')\cdot p \right)      \Bigg)\nonumber  \\
	&+ \frac{C_2(Q^2)}{\sqrt{q^2}} \Bigg( \frac{q^2}{2 \sqrt{3} \, m} \Bigg) \Pi^\mu \Bigg( \left( \epsilon^*(\vec{p}',\lambda')\cdot \epsilon(\vec{p},\lambda) \right) - \frac{m(2m' + m - \omega)}{\Omega}  \left( \epsilon(\vec{p},\lambda)\cdot p'\right) \, \left( \epsilon^*(\vec{p}',\lambda')\cdot p \right)      \Bigg) 
   \Bigg] \nonumber
\end{align}
where $\Pi^\mu = (p'+p)^\mu - \frac{m'^2 - m^2}{q^2}(p'-p)^\mu$, $\Omega \equiv (p \cdot p')^2 - m^2 m'^2$ and  $\omega = \frac{m^2 - m'^2 + q^2}{2m }$.
\end{widetext}

\begin{figure}
 \includegraphics[width=9.5cm,bb=0 0 661 439]{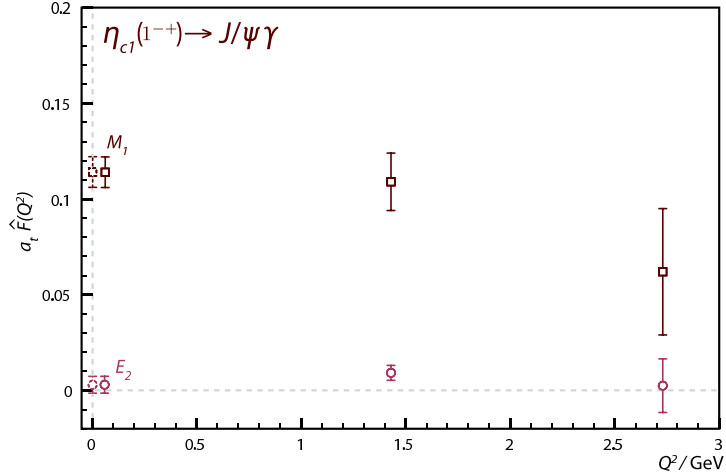}
\caption{\label{pi1rho}Magnetic dipole and electric quadrupole form-factors for the transition $\eta_{c1}(1^{-+}) \to J/\psi \gamma$.  Point at $Q^2=0.06$ GeV translated to the physical photon point $Q^2=0$.}
\end{figure}

We used a set of eight operators at the sink to produce the $T_1^{-+}$ state\footnote{In the nomenclature of \cite{Dudek:2007wv} they are smeared and unsmeared versions of $a_{0(2)}\times \nabla_{T1}, b_1\times \nabla_{T1}, \rho\times \mathbb{B}_{T1}, \rho_{(2)}\times \mathbb{B}_{T1}$. The $\mathbb{B}$ operators induce an essential gluonic component through a factor proportional to $F^{\mu\nu}$ in the continuum limit, the $\nabla$ operators reduce to covariant derivatives and hence have a factor $A^\mu$.}. Our results projected on to the ground state in $T_1^{-+}$ are shown in figure \ref{pi1rho} - the lattice data are not fitted as a function of $Q^2$ as very little extrapolation is required to associate the point at $Q^2=0.06\,\mathrm{GeV}$ with the real photon point. The value of $M_1(0)$ corresponds to a partial decay width $\Gamma(\eta_{c1} \to J/\psi \gamma) = 115(16)$ keV. Note that this is no different in scale to many measured conventional charmonium transitions.

\begin{figure}
 \includegraphics[width=9.5cm,bb=0 0 661 439]{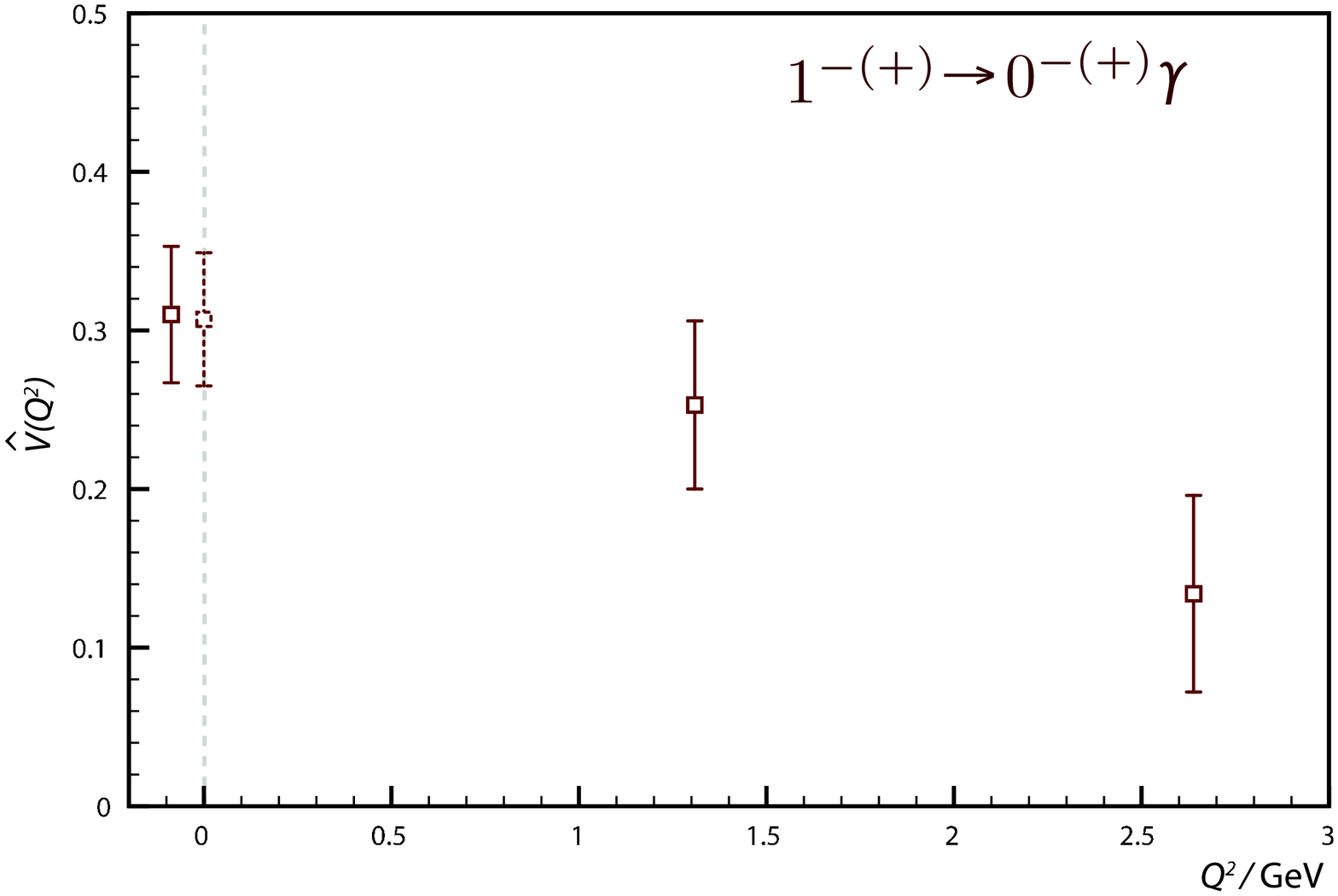}
 \includegraphics[width=9.5cm,bb=0 0 661 439]{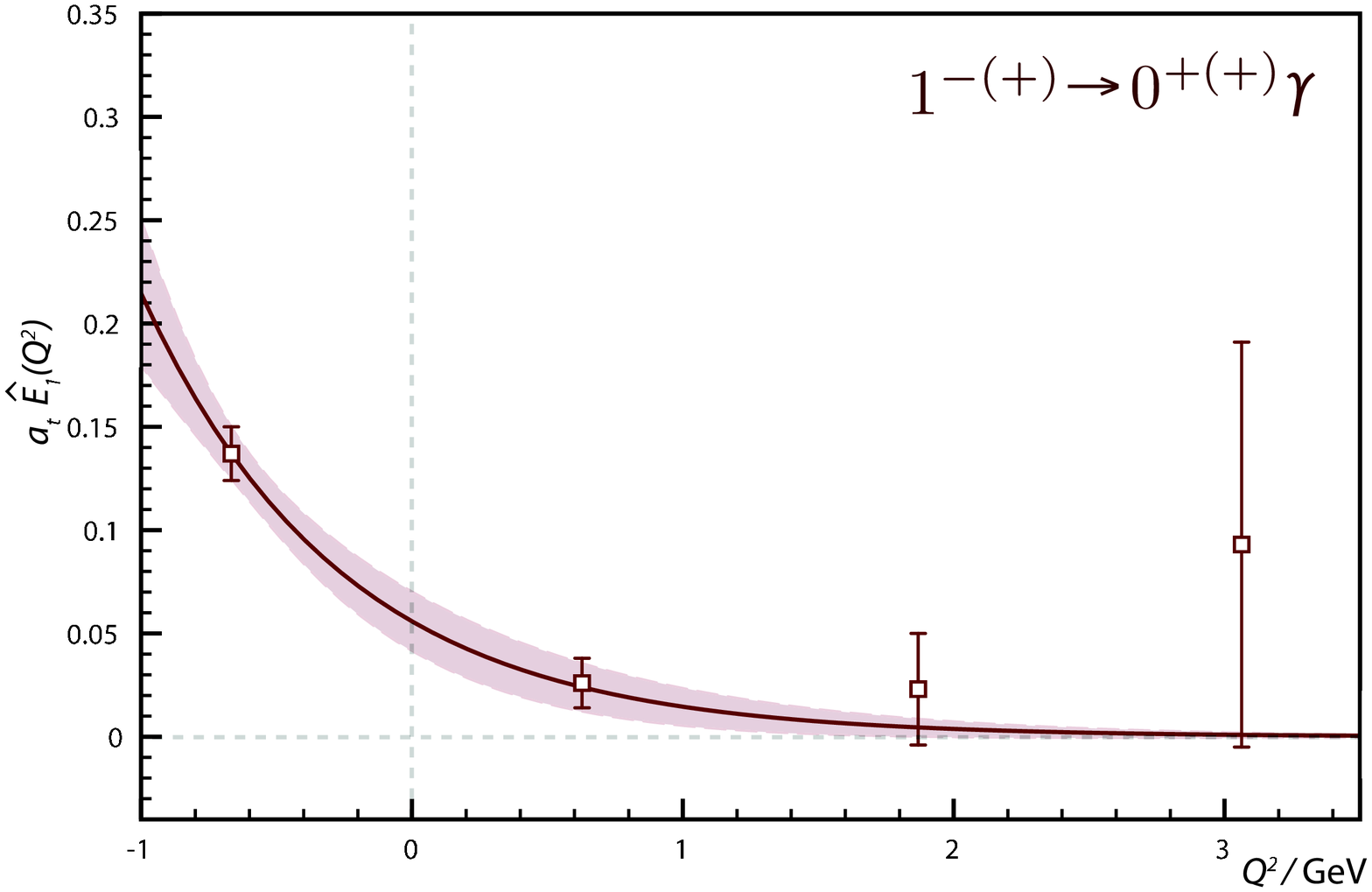}
\caption{\label{pi1_SV}(a) $1^{-(+)} \to 0^{-(+)} \gamma$ magnetic dipole transition. Linear interpolation using the two lowest-lying $Q^2$ points gives the value shown at $Q^2=0$. (b) $1^{-(+)} \to 0^{+(+)} \gamma$ electric dipole transition. Fit is with an exponential in $Q^2$.}
\end{figure}

If we also consider charge-conjugation symmetry violating decays (by coupling only to one quark) we have access to  $1^{-(+)} \to 0^{-(+)} \gamma$, $1^{-(+)} \to 0^{+(+)}\gamma$. These results can be compared to models of hybrid bound-state structure as we will do in the following section. Results are shown in figure \ref{pi1_SV}.

\begin{figure}
 \includegraphics[width=9.5cm,bb=0 0 661 439]{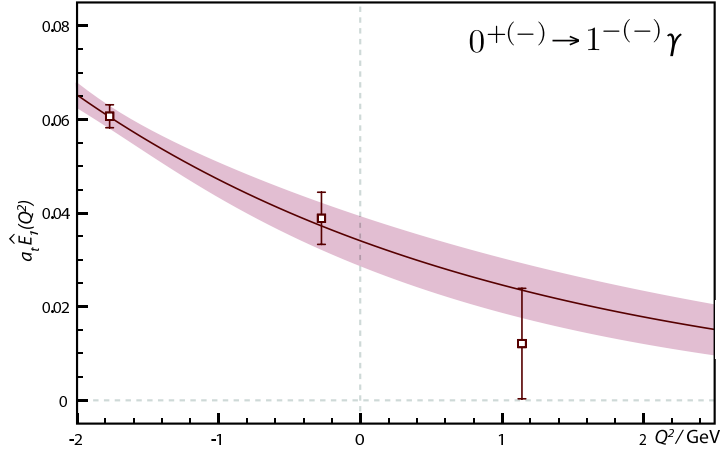}
\vspace{-8mm}
\caption{\label{b0rho}Electric dipole form-factor for the transition $0^{+(-)} \to 1^{-(-)} \gamma$.  Fit is an exponential in $Q^2$.}
\end{figure}

Another exotic $J^{PC}$ to which we have easy access is the $0^{+-}$ state which appears in the $A_1^{+-}$ channel at $4465(65)$ MeV. We have computed a charge-conjugation symmetry violating decay of the lightest state with these quantum numbers to the conventional ground state vector $1^{-(-)}$. This transition matrix element has the same decomposition as the vector-scalar in equation \ref{SV}. We used a set of three operators\footnote{quark smeared $\bar{\psi} \gamma^0 \psi$ and smeared and unsmeared $a_1\times \mathbb{B}_{A1}$} to create the $0^{+(-)}$ state; in figure \ref{b0rho} we show the electric dipole form-factor for $0^{+(-)} \to 1^{-(-)} \gamma$.

\subsection{Other assorted transitions}
\begin{figure}
 \includegraphics[width=9.5cm,bb=0 0 661 439]{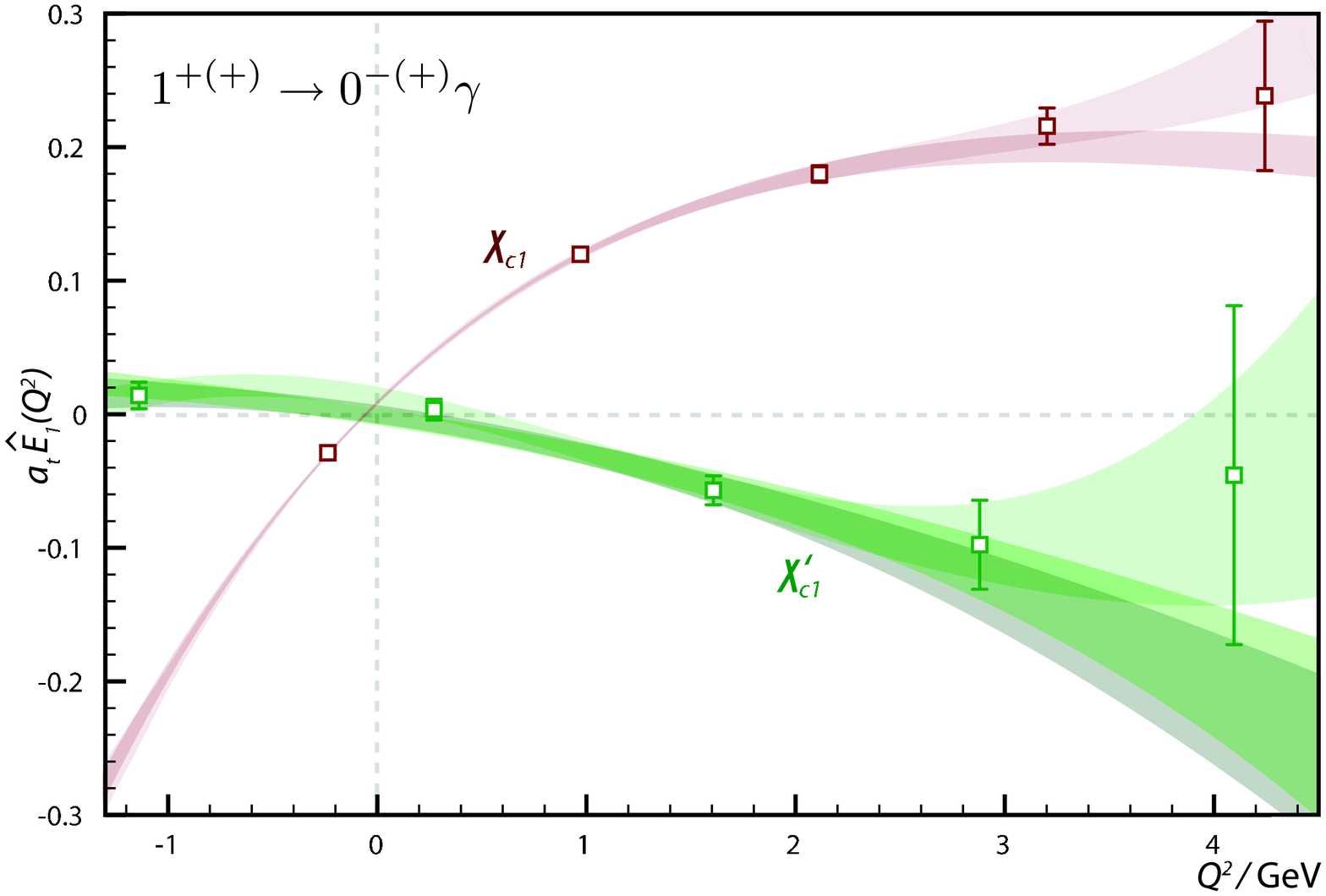}
 \includegraphics[width=9.5cm,bb=0 0 661 439]{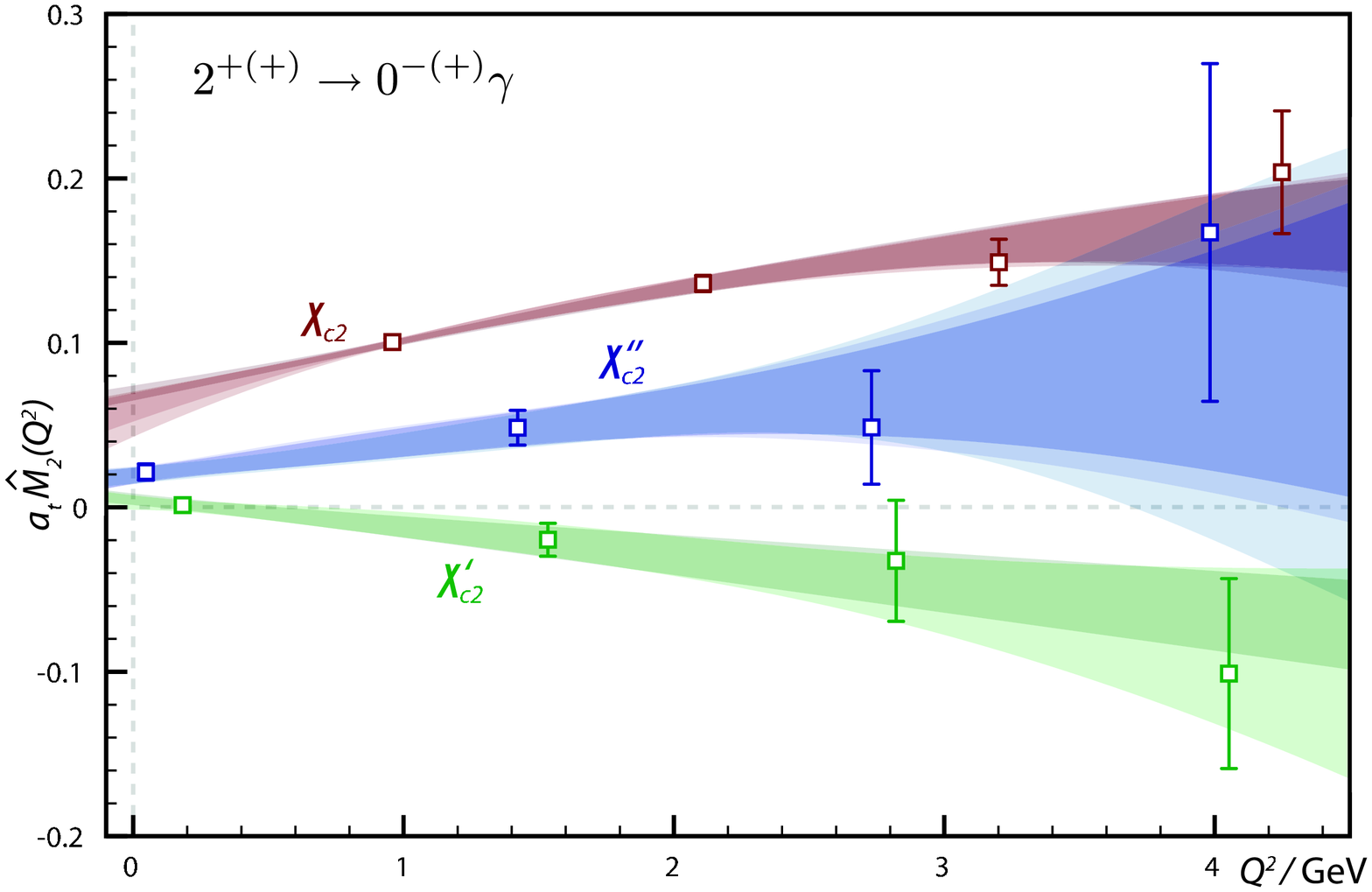}
\caption{\label{etc} (a) Electric dipole form-factor for the transition $1^{+(+)} \to 0^{-(+)} \gamma$. (b) Magnetic quadrupole form-factor for the transition $2^{+(+)} \to 0^{-(+)} \gamma$. Note that relative signs between states are not relevant.}
\end{figure}

We also computed some $C$-violating transitions which have allowed analogues in the light quark sector\footnote{namely $a_1^\pm \to \pi^\pm \gamma$ and $a_2^\pm \to \pi^\pm \gamma$.}, $1^{+(+)} \to 0^{-(+)} \gamma$ and $2^{+(+)} \to 0^{-(+)} \gamma$. The first of these has a multipole decomposition identical to eqn \ref{SV} while the second takes the form
\begin{equation}
 \langle P(\vec{p}_V) |j^\mu(0) | T(\vec{p}_T, \lambda)\rangle = M_2(Q^2)\,  \sqrt{2} \frac{m_T}{\Omega}\epsilon^{\mu\nu\rho\sigma}p_{T\rho} p_{P\sigma} \epsilon_{\nu\tau}(\vec{p}_T,\lambda) p_P^\tau \nonumber
\end{equation}
Results are shown in figure \ref{etc}. A discussion in terms of non-relativistic quark models follows in the next section. 

\section{Phenomenology}\label{phensec}

The results presented in the previous section can be compared with experimental measurements and additionally with the predictions of various models describing heavy quark bound states. A rather successful approach to describing charmonium has come by considering heavy charm quarks to be moving non-relativistically (or nearly so) in a static potential motivated by QCD. These quark potential models have been used to compute spectra, radiative transition rates, and, in many-body extensions of the theory, hadronic decay rates (e.g. \cite{Godfrey:1985xj,Eichten:1978tg, Barnes:2005pb}). In their simplest formulation, the gluonic field plays no dynamical role and one does not have exotic quantum numbered states having excited glue (hybrids). In order to consider such states one can construct models making specific assumptions about the nature of the gluonic field and its excitations, examples being the flux-tube model\cite{Isgur:1984bm}, ``constituent'' gluon models and models based upon a many-body treatment of QCD in a physical gauge\cite{Guo:2008yz}. In what follows we will address to what extent our lattice QCD results inform these models.

Firstly we should discuss the degree to which the approximations we have made in our computation introduce systematic error to our results.

\subsection{Lattice Systematics}
\begin{figure}
 \includegraphics[width=9.5cm,bb=0 0 680 480]{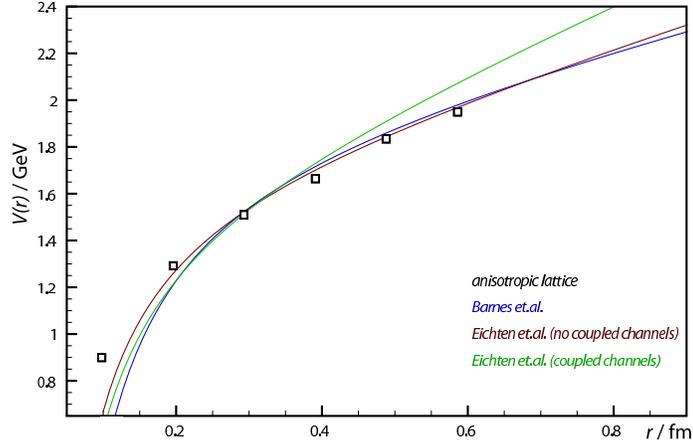}
\caption{\label{potl}Lattice static potential along with phenomenological potentials used in \protect \cite{Barnes:2005pb, Eichten:2004uh, Eichten:1979ms}.}
\end{figure}

This calculation has been performed within the quenched truncation - as such one way to view it is as a calculation of a version of QCD having just one heavy flavor of quark with the neglect of heavy flavor quark loops being justified by their large mass. In this regard this calculation is rather directly comparable to quark potential models which in their simplest form also neglect the effect of dynamical light quarks. We set the lattice scale in our calculation using the Sommer parameter which is itself related to the static quark potential and the charmonium spectrum, so here too the comparison is fairly direct. In figure \ref{potl} we show the static potential extracted on this lattice along with the phenomenological forms used in a selection of potential models.

Clearly then our calculation does not include the effect of charmonium states coupling to multi-hadron states containing light quarks, e.g. $D\bar{D}$. There have been suggestions within extensions of quark-potential models that including such physics can have a considerable effect on the spectrum \cite{Eichten:1978tg, Barnes:2007xu} and radiative transition rates\cite{Li:2007xr} of charmonium. We will attempt to address this in the discussion to follow.

In terms of direct comparison with reality, in which there are three flavors of quark lighter than charm, another failing introduced by the quenched approximation is the incorrect running of  the strong coupling owing to the beta function containing the wrong number of quark degrees of freedom. We expect this to show up in terms of inconsistent scale setting - something that has been observed previously \cite{Chen:2000ej}.

Our calculation has been performed at only one value of the lattice spacing; as is well known one expects there to be systematic shifts in quantities owing to the lattice discretization that go away as one approaches the continuum. These are suggested to be particularly serious for the discretization of heavy quarks. At a single lattice spacing we cannot accurately estimate the size of such effects, although our use of the improved Clover action and an anisotropic lattice\footnote{Some comparison of the Clover action with the Domain Wall action on these lattices is given in \cite{Dudek:2007wv}} should help a great deal in the reduction. We present in Appendix B a limited study of improving the vector current insertion in the manner described in \cite{Harada:2001ei}. In general the improvement effects are not large, suggesting that scaling to the continuum should not overwrite our results.

In the spectrum calculations using the same action on the same lattices\cite{Dudek:2007wv, Dudek:2008sz}, the possibility that the spatial volume is too small to comfortably house highly excited states was raised. 

One way to reduce the systematic error introduced by extrapolating from finite $Q^2$ to $Q^2=0$ would be to consider utilizing twisted boundary conditions on the propagator inversions  to get momenta and hence $Q^2$ values rather close to $Q^2=0$ - this has proven successful in form-factor computations\cite{Boyle:2007wg}.

Modulo the caveats that we have raised, we remain convinced that the results we present likely represent a faithful description of the pattern of physics of radiative transitions in charmonium, including the properties of gluonic hybrids. Most importantly we have clearly demonstrated that the technology of projecting three-point functions using ``ideal'' operator eigenvectors obtained in two-point function calculations works well in giving us access to excited state transitions. All the possible sources of systematic error can be addressed in future calculations using a set of sufficiently large dynamical lattices of various lattice spacings.

\subsection{Conventional state transitions and the quark-potential model}
\label{phenom}

Ref.\ \cite{Dudek:2008sz} used information extracted from two-point functions to identify excited charmonium states found in a lattice calculation with the corresponding states expected in the quark model.  The transition form-factors can also be compared to calculations in quark-potential models.  Our quenched lattice calculations offer a rather direct comparison with these potential models: in both cases loops of mesons containing light quarks have been ignored.  As discussed in Ref.\ \cite{Dudek:2006ej, Lakhina:2006vg}, the results of quark models are sensitive to approximations, such as the choice of frame, which are not an issue in lattice calculations - we will not discuss these further here.  In the following, for simplicity, we use harmonic oscillator wavefunctions and consider the rest frame of the initial meson.

Within the simplest non-relativistic model of a meson emitting a photon of momentum $\vec{q}$, we find the following transition form factors for non-radially excited states undergoing a change of orbital angular momentum of one unit:
\begin{equation}
E_1(|\vec{q}|^2) \propto \left[ 1 + r \frac{|\vec{q}|^2}{4 \beta^2} \right] \exp \left( -\frac{|\vec{q}|^2}{16 \beta^2} \right)
\end{equation}
where $\beta$ is the harmonic oscillator wavefunction parameter.  Here the $r$-values follow from the coupling of quark spin and orbital angular momentum as\footnote{note that a typographical error in \cite{Dudek:2006ej} is corrected here}
\begin{equation}
 \begin{matrix}
  r=1 & \chi_{c0}(^3P_0) \to J/\psi(^3S_1)  \gamma\\
r = 1/2 & \chi_{c1}(^3P_1) \to J/\psi(^3S_1)  \gamma\\
r = -1/2 & \chi_{c2}(^3P_2) \to J/\psi(^3S_1)  \gamma\\
r = 0 & h_c(^1P_1) \to \eta_c(^1S_0) \gamma
 \end{matrix} \nonumber
\end{equation}
Transforming from $|\vec{q}|^2$ to the invariant virtuality of the photon ($Q^2$) in the rest frame of the decaying meson this corresponds to
\begin{equation}
E_1(Q^2) \propto \left[ 1 + r \frac{Q^2}{4 \beta^2} \frac{1 + \Delta}{1 + r \delta} \right] \exp \left( -\frac{Q^2}{16 \beta^2}(1 + \Delta) \right)
\label{equ:FormFac} 
\end{equation}
with $\Delta \equiv \frac{m_f^2 - m_i^2}{2 m_i^2}$ and $\delta \equiv
\frac{(m_f^2-m_i^2)^2}{16 m_i^2 \beta^2}$. Note that this is of the general form given in eqn. \ref{poly_times_exp}.

In the same model, we have for transitions involving both a change in orbital angular momentum and a quark spin-flip, the expression
\begin{equation}
E_1(|\vec{q}|^2) \propto \frac{|\vec{q}|^2}{\beta^2} \exp\left(-\frac{|\vec{q}|^2}{16 \beta^2}\right).\label{supp}
\end{equation}
This applies to the charge-conjugation forbidden decays $1^{+(+)}(^3P_1) \to 0^{-(+)}(^1S_0) \gamma$ and $1^{+(-)}(^1P_1) \to 1^{-(-)}(^3S_1) \gamma$.  In addition, this same form is predicted for the $M_2$ form-factor in the transitions $\chi_{c1,2}(^3P_{1,2}) \to J/\psi(^3S_1) \gamma$ and the charge-conjugation forbidden $2^{+(+)}(^3P_2) \to 0^{-(+)}(^1S_0) \gamma$  and $1^{+(-)}(^1P_1) \to 1^{-(-)}(^3S_1) \gamma$.  In terms of $Q^2$ this gives
\begin{equation}
E_1(Q^2) \propto \left[ \delta + \frac{Q^2}{4 \beta^2} (1 + \Delta) \right] \exp\left(-\frac{Q^2}{16 \beta^2}(1 + \Delta) \right),
\label{equ:FormFacSuppressed}
\end{equation}
which we note is also of the form of eqn \ref{poly_times_exp}.

Within this model we find that $E_3(Q^2) = 0$ for $\chi_{c2}(^3P_2) \to J/\psi(^3S_1) \gamma$.  In fact this result follows from
the general assumption of a \emph{single quark
  transition}\cite{Olsson:1984zm} and does not depend critically upon the details
of the model.  For a transition $J \rightarrow J'$ (where $J$ and $J'$
are respectively the initial and final meson spins) the allowed $k$
(for $E_k$ or $M_k$ depending on $k$ and the parity change) are in
general $|J-J'| \leq k \leq J+J'$.  However, if only a single quark is
involved in the transition there is a further restriction on the
values allowed.  If the interacting quark in the initial (final) meson
has total angular momentum $j$ ($j'$) then $|j-j'| \leq k \leq j+j'$.  Hence in general for an $n\, ^3P_2 \rightarrow
n'\,^3S_1$ transition the allowed $k$ are $k=1,2,3$.  However in the
single quark transition assumption we have $j=3/2 \rightarrow j'=1/2$ and $k$ is
restricted to $k=1,2$.  Note that it \emph{is} possible to have $E_3
\neq 0$ if the tensor meson is a $^3F_2$ ($j'=5/2$) state, in which
case $k=2,3$ and $E_1=0$.  It is also possible to have $E_3 \neq 0$ in
transitions involving hybrid, multiquark or molecular mesons where
there are other degrees-of-freedom able to carry some angular momentum.

\subsubsection{ $\psi - \eta_c$ transitions}

Our result for $\Gamma(J/\psi \to \eta_c \gamma)$ is in reasonable agreement with the new experimental value from CLEO-c \cite{Mitchell:2008fb},  the potential model calculations of  \cite{Eichten:2002qv, Barnes:2005pb}, the pNRQCD result of \cite{Brambilla:2005zw} and the QCD sum-rule estimate of \cite{Beilin:1983ht}.  Similarly our result for the ``hindered'' $M_1$ transition $\Gamma(\psi' \to \eta_c \gamma)$ is consistent with the experimental values\cite{Amsler:2008zzb, Mitchell:2008fb}  and Eichten et. al.\ \cite{Eichten:2002qv}, but significantly smaller than the values in Ref.\ \cite{Barnes:2005pb} which depend sensitively on the overlap of orthogonal wavefunctions with a small recoil factor breaking the orthogonality.

Li and Zhao\cite{Li:2007xr} study the possibility of hadronic loop contributions in
$J/\psi$ and $\psi'$ radiative decays to $\eta_c$ and $\eta_c'$.  They
find the addition of loop contributions involving pairs of $D$ and $D^*$ mesons can bring the model values from Ref.\
\cite{Barnes:2005pb} in to line with experiment.  Most significantly,
they find large loop contributions, of order $10$ keV, to the $\psi' \rightarrow \eta_c
\gamma$ transition.  A cancellation between the
quark model contribution of order $10\,\mathrm{keV}$ (from Ref.\
\cite{Barnes:2005pb}) and these loop contributions results in a small
transition width of $\sim 1\ \text{keV}$.  This appears to present a
problem with respect to our result, which being quenched does not include
loop contributions, since there is no room for a large loop contribution
when our $\Gamma(\psi' \rightarrow \eta_c \gamma)$ is compared to the
experimental value.  This might indicate that the loop contributions of Ref.\ \cite{Li:2007xr} are overestimated and we have identified one possible source for this in the coupling $g_{D^* D \gamma}$ used. The partial decay width $\Gamma(D^{*0} \rightarrow
D^0 \gamma) \sim 800\ \text{keV}$ assumed in Ref.\ \cite{Li:2007xr}
appears to be rather large for an $M_1$ radiative transition and, with
the branching ratio from the PDG\cite{Amsler:2008zzb}, implies a $D^{*0}$ total
width of $2.1\ \text{MeV}$ which is at the upper limit allowed by the PDG. For comparison, Close and
Swanson\cite{Close:2005se}, using a non-relativistic quark model predict $\Gamma(D^{*0} \rightarrow D^0 \gamma) = 32\ \text{keV}$.  The corresponding decrease in the loop amplitudes would remove the problem in comparison with our result.  In addition, there are considerable uncertainties arising from the estimated hadronic coupling constants in the loop contributions which are not discussed in Ref.\ \cite{Li:2007xr}.  These uncertainties may be particularly important because of the delicate nature of the cancellation.

We have also been able to extract a signal for the $\psi'' \to \eta_c \gamma$ which in the standard interpretation of the $\psi''$ would be a $1\,^3D_1 \to \,^1S_0$ transition. In a harmonic oscillator wavefunction basis the form-factor has the same leading $|\vec{q}|^3$ behaviour as $2\,^3S_1 \to \, ^1S_0 \gamma$  indicating that it should suffer ``hindered'' suppression just like the $\psi' \to \eta_c \gamma$ relative to $J/\psi \to \eta_c \gamma$. The lattice data for the amplitude appears to be in line with this, within a large statistical uncertainty. Discussion of the next excited vector state will follow in section \ref{exotic_discussion}.

\subsubsection{$\chi$-$\psi$ transitions}

The $\chi_{cJ} \to J/\psi \gamma$ partial widths we obtain are dominated by the electric dipole component and compare favorably with the experimental values. The largest discrepancy is in the statistically precise $\chi_{c0}$ transition. The lattice systematic errors discussed above may be to blame for this, without further calculation we cannot determine this definitively, but we note that \cite{Eichten:1979ms} found that approximately ``unquenching'' the quark model for this transition did not induce a large change in the rate. In Appendix B we consider the effect of using an improved vector current and we find that the $\chi_{c0} \to J/\psi \gamma$ transition is one case where a statistically significant change with respect to the local current does occur. This might indicate that this transition is particular sensitive to scaling in the lattice spacing.

The fitted values of $\beta$ in eqn \ref{poly_times_exp} are in rough agreement for all of $\chi_{c0,1,2}$ as one might expect for states that in the potential models differ only by small spin-orbit effects. The fitted values of $\lambda$ deviate somewhat from the quark-model expected $1:\tfrac{1}{2}:-\tfrac{1}{2}$ ratio for $\chi_{c0,1,2}$, being $1:0.62(26):-0.34(2)$. Given the approximations and frame-dependence inherent in the quark model even such rough agreement is surprising. 

In the $\chi_{c1,2} \to J/\psi \gamma$ transitions, the hierarchy of multipoles expected in the quark model ($|E_1(0)| > |M_2(0)| \gg |E_3(0)|$) is observed.  The precise degree of suppression of $M_2$ compared with $E_1$ is not easy to compute in non-relativistic quark potential models.  However, the single quark transition assumption predicts that $E_3(Q^2)=0$ independent of the frame and details of the potential and the lattice data do agree with this. Our extracted values of $M_2(0)$ or the ratio $M_2(0)/E_1(0)$ all depend upon theoretically undetermined extrapolation in $Q^2$ and so the apparent disagreement with the very small values found in experiment is not yet overly concerning.

The electric dipole couplings for $\psi' \to \chi_{c0} \gamma$ and $\psi'' \to \chi_{c0} \gamma$ are, within reasonably large statistical errors, in agreement with experimental values and with the ``GI'' values tabulated in \cite{Barnes:2005pb}\footnote{But note the considerable dependence on the details of the quark-model given in that paper's ``NR'' results}. The estimates of $D$-meson loop effects from \cite{Eichten:1979ms} are comparable to the level of our statistical uncertainty.

For the transition $\chi_{c1}' \to J/\psi \gamma$ we extract a partial width of $21(12)$ keV which is in reasonable agreement with the quark-model estimates of \cite{Barnes:2005pb}, being $14$ or $71$ keV depending upon model details. Since in our study this state has a likely interpretation as the $2\,^3P_1$ state of the quark model, without any $D\bar{D}^*$ effects included, this can act as a benchmark for models that consider the experimental $X(3872)$ state as being $1^{++}$.

The large $E_3(0)$ in the lattice calculation of the $\chi_{c2}' \to
J/\psi \gamma$ transition is in stark disagreement with the general
result that $E_3=0$ for a $^3P_2 \rightarrow \, ^3S_1$ transition.  As
discussed above, this does not appear to be a lattice artifact and so
another interpretation must be found.  Possible explanations include
that the $\chi_{c2}'$ state is a hybrid where the gluonic field carries spin,
but here we concentrate on the most conservative interpretation, that
$\chi_{c2}'$ is a conventional $^3F_2$ state.

An $F$-wave tensor state ($^3F_2$) is expected to have $E_1(Q^2) = 0$
in general from the single quark transition assumptions outlined earlier.  In addition, a quark model calculation shows that $M_2(0)$ is highly suppressed:
\begin{equation}
M_2(|\vec{q}|^2) \propto \frac{|\vec{q}|^4}{\beta^4} \exp\left(-\frac{|\vec{q}|^2}{16 \beta^2}\right) \nonumber
\end{equation}
whereas $E_3$ is less suppressed:
\begin{equation}
E_3(|\vec{q}|^2) \propto \frac{|\vec{q}|^2}{\beta^2}  \exp\left(-\frac{|\vec{q}|^2}{16 \beta^2}\right) \nonumber
\end{equation}
with the same leading $|\vec{q}|^2$ dependence as the $M_2$ form factor
in a $^3P_2 \rightarrow \,^3S_1$ transition
(Eqn. \ref{equ:FormFacSuppressed}).  This pattern that $|E_3(0)| >
|M_2(0)| \gg |E_1(0)|$ is consistent with the lattice results for this
$\chi_{c2}' \to J/\psi \gamma$ transition.  These lattice results
support the interpretation of this $\chi_{c2}'$ as the lightest
$^3F_2$ state having a partial decay width to $J/\psi \gamma$ of $20(13)$ keV. We are unaware of any model calculations of this transition rate. 

The $\chi_{c2}'' \to J/\psi \gamma$ transition appears to have reverted to the $|E_1(0)| > |M_2(0)| \gg |E_3(0)|$ hierarchy expected for a $^3P_2$ state and we propose that this state is the $2\,^3P_2$.  There is of course the possibility that the $P$ and $F$ states are mixed, but this mixing cannot be anything like maximal given a large $E_3$ appears in only one case and a large $E_1$ only in the other case. Within a non-relativistic quark model, the $^3P_2$ and $^3F_2$ states can mix via any tensor potential in the Hamiltonian.  However in most models this term is suppressed by the inverse charm mass squared and
so is small. The tensor term, like the hyperfine interaction, is a short distance effect and so may not be reliable in these quenched lattice calculations.  Because this term is suppressed anyway, the errors introduced from this should be small. Another possible source of mixing which we lack in the quenched theory comes through $D\bar{D}$ meson loops - such effects are discussed in \cite{Barnes:2007xu, Close:2009ii}.

The partial decay width $\Gamma(\chi_{c2}'' \to J/\psi \gamma)$ extracted from the lattice data is $88(13)$ keV which is comparable with the $53$ or $81$ keV computed by \cite{Barnes:2005pb}. We note that the interpretation of excited states in this channel as $F$ and $P$ waves was hinted at in the two-point
function analysis of \cite{Dudek:2008sz} where states compatible with being $^3F_2$ and $2\,^3P_2$ were found to be nearly degenerate. Within the quark potential models used in \cite{Barnes:2005pb}, the $F$ wave state is expected $60-100$ MeV heavier than the $P$ wave. The degeneracy we found may be an artifact of ``squeezing'' these spatially larger states into a $1.2$ fm box.

A discussion of the excited $\chi_{c2}$ states is timely given the observation of a candidate state at $3929(5)$ MeV by Belle in $\gamma \gamma \to D\bar{D}$ which has been associated with the $2\,^3P_2$ state\cite{Uehara:2005qd}. We lack reliable theoretical estimates of the $\gamma \gamma$ widths of $^3F_2$ and $2\,^3P_2$ states - this calls for an extension of the work done in \cite{Dudek:2006ut} to consider excited states which could use very similar operator projection technology to that discussed in this paper. We note here that an optimist viewing figure 2 of \cite{Uehara:2005qd} might hope that the statistically insignificant excess at $4080$ MeV could, with increased statistics, become a signal for the other state in the $2\,^3P_2$ / $1\,^3F_2$ pair.

The $J^{+(+)} \to 0^{-(+)} \gamma$ transitions shown in Figure \ref{etc} are consistent with the quark model predictions. The electric dipole amplitudes in $1^{+(+)} \to 0^{-(+)} \gamma$ are much smaller than other electric dipole amplitudes since they involve both $\Delta L=1$ and a spin-flip. For the magnetic quadrupole amplitudes in $2^{+(+)} \to 0^{-(+)} \gamma$, the $^3P_2$ amplitude behaves like Eqn \ref{supp} while the $^3F_2$ amplitude is suppressed by a further two powers of $|\vec{q}|$ which suggests yet again the assignment $\chi_{c2} = 1\,^3P_2$, $\chi_{c2}' \approx 1\,^3F_2$ and $\chi_{c2}'' \approx 2\,^3P_2$.

\subsection{Exotic \& crypto-exotic transitions}\label{exotic_discussion}

The only charge-conjugation allowed transition involving an exotic we compute, $\eta_{c1} \to J/\psi \gamma$, has a rather large partial decay width $115(16)$ MeV which is dominantly through a magnetic dipole transition. Even accounting for the large phase space this is very large on the usual scale of magnetic dipole transitions. Conventional $c\bar{c}$ states can only have magnetic dipole transitions if there is a quark spin-flip and this is suppressed by the large charm-quark mass. In a hybrid meson the extra gluonic degree-of-freedom allows an $M_1$ transition \emph{without spin-flip}. This can be seen explicitly within the flux-tube model, where in an $M_1$ transition between a conventional $L=0$ meson and a hybrid meson the tube absorbs the angular momentum\footnote{Details are presented in the unpublished DPhil. thesis of JJD where the following partial widths were obtained: $\Gamma(\eta_{c1} \to J/\psi \gamma) \approx \Gamma(Y_{\mathrm{hyb}} \to \eta_c \gamma) \approx 30 \to 60 \,\mathrm{keV}$ }.

The two-point function analysis of \cite{Dudek:2008sz} suggested that the excited vector state we have called $Y_{\mathrm{hyb.}}$ (having mass around $4.4$ GeV) is a crypto-exotic hybrid meson, having conventional $J^{PC}=1^{--}$ but internally an excited gluonic field. That analysis preferred the quarks to be in a spin-singlet in this state. In that case we would expect a large non spin-flip $M_1$ transition $Y \to \eta_c \gamma$ which is precisely what is seen in our lattice data. Here $\Gamma(Y \to \eta_c \gamma) = 42(18)\,\mathrm{keV}$ which is considerably larger than any other vector to pseudoscalar transition. Within the flux-tube model, the non-exotic $Y(1^{--}_\mathrm{hyb.})$ and the exotic $\eta_{c1}(1^{-+}_\mathrm{hyb})$ (which we find at a mass $\sim 4.3$ GeV) differ only in being quark spin singlet and triplet respectively and $\Gamma(Y\to \eta_c \gamma) \approx \Gamma(\eta_{c1} \to J/\psi \gamma)$. The lattice data does not strongly disagree with this. It would be very interesting to see analogous calculations in other models of excited glue to see if this is a general result or one peculiar to the flux-tube model.

We also consider charge-conjugation violating electric dipole transitions involving exotic mesons, $1^{-(+)} \to 0^{+(+)} \gamma$ and $0^{+(-)} \to 1^{-(-)} \gamma$, where for both we found non-zero signals. Roughly speaking, one can gauge the cost of exciting the gluonic field versus exciting conventional orbital angular momentum by comparing the electric dipole transitions $1^{-(+)} \leftrightarrow 0^{+(+)}$ having $E_1(0) = 0.06(1)$ and $1^{-(-)} \leftrightarrow 0^{+(+)}$ having $E_1(0) = 0.127(2)$. These numbers are clearly of the same order, a result that is also true in the flux-tube model\cite{Close:2003fz, Close:2003ae} - whether this is true in other models has not to our knowledge yet been tested. The transition $0^{+(-)} \to 1^{-(-)}$ where $E_1(0) \sim 0.04(1)$ suggests that $L_{q\bar{q}}=0$ transitions to exotic hybrids are also unsuppressed.

Further evidence for the quark spin triplet nature of the $\eta_{c1}(1^{-+}_\mathrm{hyb})$ and the quark spin singlet nature of the $Y(1^{--}_\mathrm{hyb.})$ comes from the fact that $E_1(0)$ for $1^{-(+)} \to 0^{+(+)} \gamma$ is $0.06(1)$ while for  $1^{-(-)}_\mathrm{hyb} \to 0^{+(+)} \gamma$ it is consistent with zero. The first of these then is $S_{q\bar{q}}=1 \to S_{q\bar{q}}=1 $ while the second requires a spin-flip, expected to be suppressed by the heavy-quark mass. Note that the magnetic dipole transition $1^{-(+)} \to 0^{-(+)} \gamma$ in figure \ref{pi1_SV} is at the scale of other ``hindered'' $M1$ transitions, consistent with requiring both gluonic excitation and quark spin-flip. We note that the spin singlet nature of our $Y$ state does not make it a good candidate for the experimental $Y(4260)$ state whose potentially large decay rate into $\pi \pi J/\psi$ would tend to suggest dominance of spin-triplet.

Here we briefly comment that if this heavy-quark hybrid physics is any guide to the behaviour of light-quark hybrid systems, we should expect the GlueX experiment to copiously photoproduce hybrid mesons off the meson cloud around a baryonic target.

\section{Conclusions}\label{conc}

We have demonstrated that the ``ideal'' excited state operators within a basis of operators can be used to successfully extract excited state transitions from three-point correlators. Using this technique we have carried out an extensive survey of radiative transitions in charmonium with detailed consideration of the phenomenology suggested by the results.

Notably we have performed the first lattice QCD calculation of the exotic $\eta_{c1}$ radiative decay and found a large $\Gamma(\eta_{c1} \to J/\psi \gamma) = 115(16)\ \text{keV}$.  We found statistically significant electric dipole and magnetic quadrupole form factors in $\chi_{c2} \to J/\psi \gamma$, calculated for the first time in this framework, and have studied excited $\chi_{c1,2}$ transitions. Our results for the excited $\chi_{c2}$ states suggest that there could be a radially excited $2\,^3P_2$ state and a $1\,^3F_2$ state rather close in mass. This signal followed from a clear observation of dominance of $E_1$ over $E_3$ in one case and $E_3$ over $E_1$ in the other and matches the general expectations of potential models. This is relevant given the recent observation of a candidate $2^{++}$ excited state in $\gamma \gamma \to D \bar{D}$\cite{Uehara:2005qd}.

Our calculation of magnetic dipole transition widths, such as $\psi \to \eta_c \gamma$ and hindered excited state transitions reflected the expected suppression of the excited state transitions. We note that, modulo lattice systematic effects which can be reduced, our method is not troubled by the uncontrolled approximations and model dependence inherent in model calculations of such suppressed transitions. In light of this we were able to make statements regarding the possible influence of closed channel $D$-meson loops on the hindered transitions, determining that they cannot be as large as suggested in certain studies. We identified a putative non-exotic hybrid state having a large magnetic dipole transition amplitude to $\eta_c$ - the possibility that this reflects the non-spin-flip magnetic dipole excitation allowed within the flux-tube model (and likely within other models having more than $c\bar{c}$ content) was raised.

While the current numerical results may be affected by lattice systematic errors, future calculations using the now proven methodology can address these shortcomings by using dynamical lattices of sufficient size having a number of different lattice spacings. At some point after the introduction of dynamical lattices, the difficulty of dealing with resonant states embedded in a multi-meson continuum will have to be addressed. 

We have demonstrated that our results are in agreement with general predictions of quark potential models but we are able to go beyond this to make statements about states in which there is an excited gluonic field. These results are appropriate for comparison with models proposing particular forms for the gluonic excitation.

The great advantage of this development within lattice QCD as compared to models relying upon the non-relativistic motion of quarks, is its immediate applicability to the light quark systems. Future efforts will consider photocouplings of light quark mesons, and in particular exotics, as these are of central important in the production rates for the GlueX project. If the large couplings we find for the $1^{-+}$ state with heavy quarks persist into the light quark sector this will confirm the intuition and model results used to motivate the GlueX production process.

\begin{acknowledgments}
We acknowledge fruitful discussions with our colleagues David Richards and Nilmani Mathur. We thank Qiang Zhao for his communications regarding his research on hadron loops and Matt Shepherd and Ryan Mitchell for their communications regarding CLEO-c.
Notice: Authored by Jefferson Science Associates, LLC under U.S. DOE Contract No. DE-AC05-06OR23177. The U.S. Government retains a non-exclusive, paid-up, irrevocable, world-wide license to publish or reproduce this manuscript for U.S. Government purposes. Computations were performed on clusters at Jefferson Laboratory as part of the USQCD collaboration.
\end{acknowledgments}

\appendix

\section{Derivative operators at zero and non-zero momentum}
In this appendix we give a few details of the spin-structure we deal with when considering two- and three- point correlators. We take the example of the $T1^{--}$ ``vector'' operators considered at $\vec{p}=(000)$ and $\vec{p}=(100)$. Using the decompositions in the appendix of \cite{Dudek:2007wv}, we find that at zero momentum all the operators we have used have the following behaviour (at least in the continuum limit),
\begin{equation}
 	\langle 1^{--}(\vec{p}=\vec{0}, \lambda) | {\cal O}_i | 0\rangle \propto \epsilon^*_i(\vec{0}, \lambda),
\end{equation}
and there are no non-zero overlaps onto particle of any other $J^{PC}$ (except those lattice artifacts that appear suppressed by powers of $a$).
The three-point correlator then involves a sum $\sum_\lambda \epsilon^*_i(\vec{0}, \lambda) \epsilon_j(\vec{0}, \lambda) = -g_{ij}$.

At non-zero momentum things are not so simple, with one unit of lattice momentum (e.g. $\vec{p}=(100)$) we have the following non-zero overlaps in the continuum limit,
\begin{align}
 	\langle 1^{--}(\vec{p}, \lambda) | {\cal O}_i | 0\rangle &\sim \epsilon^*_i(\vec{p}, \lambda),\, p_j \epsilon^*_j(\vec{p}, \lambda)  p_i \nonumber\\
	\langle 1^{+-}(\vec{p}, \lambda) | {\cal O}_i |0\rangle &\sim \epsilon_{ijk} \epsilon^*_j(\vec{p}, \lambda) p_k , \nonumber \\
	\langle 0^{+-}(\vec{p}) | {\cal O}_i | 0 \rangle &\sim p_i ,\nonumber \\
	\langle 2^{+-}(\vec{p})|{\cal O}_i |0 \rangle &\sim p_j \epsilon^*_{ij}(\vec{p}, \lambda) ,\, p_i \epsilon^*_{00}(\vec{p}, \lambda)  .
\end{align}
These forms are derived following the decompositions given in Appendix A of \cite{Dudek:2007wv}\footnote{correcting some minor typographical errors therein}.

\begin{figure*}
\includegraphics[width=16cm,bb=0 0 648 326]{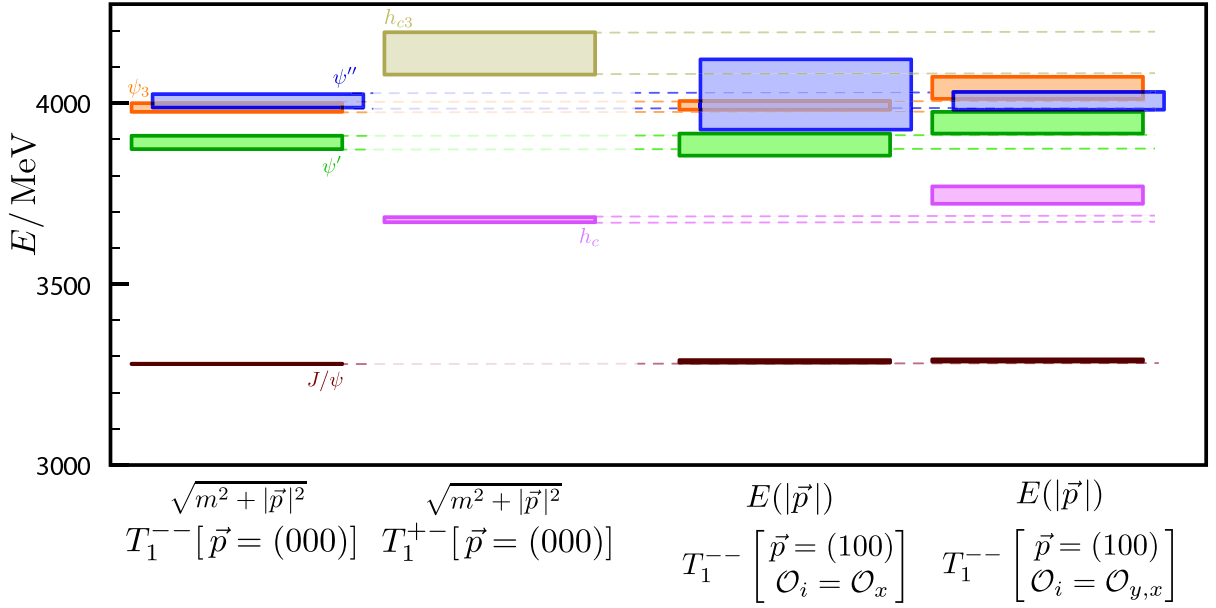}
\caption{\label{p100_levels}Low lying spectrum extracted from two-point correlators. State assignment (color-coding) follows from the consideration of overlap factor as described in the text.}
\end{figure*}

So here we have additional contributions from $(0,1,2)^{+-}$ particles - the $(0,2)^{+-}$ contributions we can neglect as these states are very heavy, but the $1^{+-}$ we must worry about. We can see the entry of such particles in the two-point function spectral analysis; in figure \ref{p100_levels} we show the spectrum at $\vec{p}=(100)$ with either $i=x$, where $1^{+-}$ does not contribute, or with $i=y,z$ where $1^{+-}$ can contribute. Also shown are the $T_1^{--}$ and $T_1^{+-}$ spectra extracted at $\vec{p}=(000)$ extrapolated to $\vec{p}=(100)$ using the continuum dispersion relation $E = \sqrt{m^2 + |\vec{p}|^2}$. We see the entry of the lightest $1^{+-}$ state into the $T_1^{--}(\vec{p}=(100))$ spectrum in the case $i=y,z$ but not in the case $i=x$ as expected. We can explicitly exclude this state from our three-point analysis by not projecting onto the eigenvector belonging to this level. In the results presented in figures \ref{SVfig},\ref{PVfig} we separately consider the cases $i=x$ and $i =y,z$ shown by the triangles and the circles. Note that we did not simply rely upon energy matching to determine the levels - a more precise mapping between $\vec{p}=(000)$ levels and $\vec{p}=(100)$ comes from considering the overlaps, $Z$. As a concrete example consider the matrix elements $ \langle 1^{--}(\vec{p},\lambda) | \bar{\psi} \gamma_i \psi | 0 \rangle = Z \epsilon_i^*(\vec{p},\lambda)$ where $Z$ is proportional to the vector decay constant and where, for unsmeared quark fields in the continuum limit, $Z$ should be independent of momentum $\vec{p}$. As can be clearly seen in Table \ref{Ztab}, one can identify the $J/\psi$ state with the lightest in each case, while the $\psi'$ state is the next lightest in two of the cases, but is the third state in the case where the $1^{+-}$ can contribute. We considered the $Z$ values for the entire set of operators when making the state assignments at finite momentum. This approach becomes increasingly more cumbersome as the momentum increases so we have not considered it any further.

\begin{table}
 \begin{tabular}{c|ccc}
  level & $\vec{p}=(000)$ & $\begin{matrix}\vec{p}=(100) \\ {\cal O}_i = {\cal O}_x \end{matrix}$  & $\begin{matrix}\vec{p}=(100) \\ {\cal O}_i = {\cal O}_{y,z} \end{matrix}$\\
\hline
0 & $\mathbf{0.163(1)}$ & $\mathbf{0.173(1)}$ & $\mathbf{0.166(1)}$  \\
1 & $\mathit{0.190(6)}$ & $\mathit{0.205(10)}$ & $0.04(1)$ \\
2 & $0.05(25)$ & $0.01(3)$ & $\mathit{0.201(12)}$
 \end{tabular} 
\caption{$Z$ values in $ \langle 1^{--}(\vec{p},\lambda) | \bar{\psi} \gamma_i \psi | 0 \rangle = Z \epsilon_i^*(\vec{p},\lambda)$. State proposed to be $J/\psi$ in bold, state proposed to be $\psi'$ in italics. \label{Ztab} }
\end{table}

\section{Improvement of the vector current}

In \cite{Harada:2001ei} the improvement of the vector current to ${\cal O}(ma)$ in a manner compatible with the improvement in the anisotropic Clover action was presented. Therein the analysis was for heavy-light currents and the renormalization constants were determined perturbatively. Here we are interested in heavy-heavy currents and we shall determine $Z_V$ non-perturbatively. Nevertheless we can consider modifying our local current to include extra terms suggested by the improvement scheme to investigate any change in the form-factor values. This can be considered to give a crude estimate of how much we might anticipate scaling to the continuum to affect our results.

Our Clover action uses $r=1$ and as such our local vector current is not automatically improved - at tree level the improved current is given by
$\bar{\Psi} \gamma_\mu \Psi$ where $\Psi(x) \propto \left(1 + a_s d_1 \gamma_j \overrightarrow{D}_j \right)\psi(x)$ is the ``rotated'' field used in the construction of the improved action. The improvement parameter $d_1 = \tfrac{1}{4}(1- \xi r) + {\cal O}(m_0 a_t) \approx -0.5$ for renormalised anisotropy $\xi=3.0$. The quark mass parameter that appears in our Clover action, $m_0 a_t$ has the value $0.0401$ which is clearly small, while the mass in spatial lattice units is three times as large, but still might be argued to be small.

We have attempted to construct the improved current by using the tree-level equation of motion
\begin{equation}
 	\left(a_t\gamma_4 \nabla_4 + a_s\tfrac{\nu}{\xi_0} \gamma_j \nabla_j + m_0 a_t\right)\psi = 0
\end{equation}
to eliminate the derivative in the expansion to ${\cal O}(a_s)$ of the improved current $\bar{\Psi} \gamma_\mu \Psi$. This yields the result
\begin{align}
 	\bar{\Psi} \gamma_4 \Psi &\propto \bar{\psi} \gamma_4 \psi - d_1 a_s \partial_j \big( \bar{\psi} \sigma_{j4} \psi \big) \\
	\bar{\Psi} \gamma_i \Psi &\propto \big(1 - 2 m_0 a_t \tfrac{\xi_0}{\nu} \big) \bar{\psi} \gamma_i \psi - d_1\tfrac{\xi_0}{\nu} a_t \partial_4 \big( \bar{\psi} \sigma_{i4} \psi \big)
\end{align}
where a common proportionality factor is ignored given that we determine $Z_V$ non-perturbatively using the pseudoscalar form-factor at $Q^2=0$. At $\vec{q}=(000)$ the second term does not contribute and we would expect $\tfrac{Z_{V(s)}}{Z_{V(t)}} = 1 - 2 m_0 a_t \tfrac{\xi_0}{\nu} d_1 \approx 1.11$ for the parameters in our action. Note that in section \ref{ZV} we found $1.11(1)$ for this ratio using a non-perturbative extraction.

\begin{figure}
\includegraphics[width=9.5cm,bb=0 0 680 480]{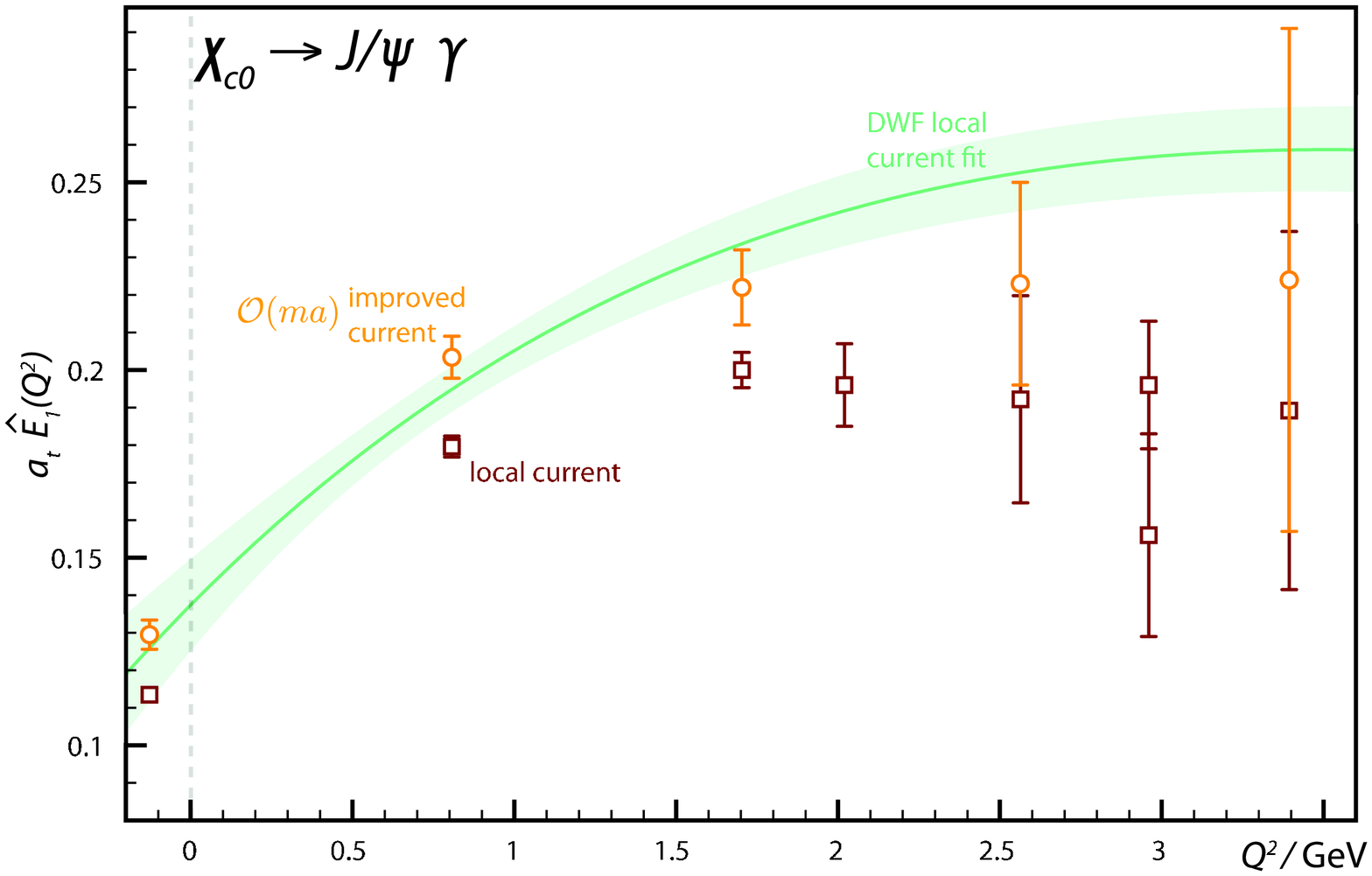}
\includegraphics[width=9.5cm,bb=0 0 680 480]{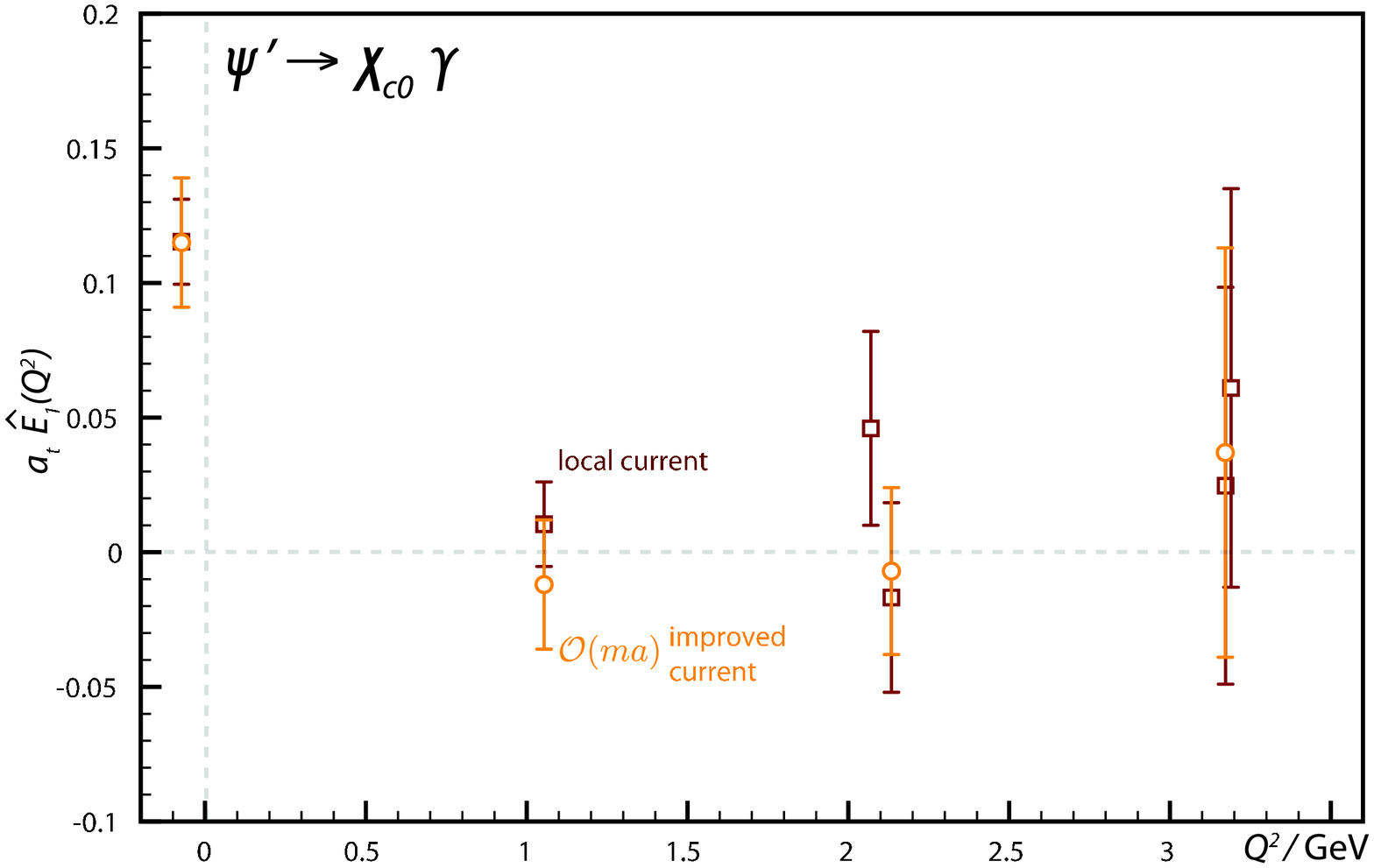}
\vspace{-8mm}
\caption{\label{SVimp}Scalar-Vector $E_1$ transition form-factors using local and ${\cal O}(ma)$ improved current with anisotropic Clover quarks and the fit from a study using the local current with DWF fermions on the same anisotropic lattices. (a) $\chi_{c0} \to J/\psi \gamma$, (b) $\psi' \to \chi_{c0} \gamma$.  }
\end{figure}

\begin{figure}
\includegraphics[width=9.5cm,bb=0 0 680 480]{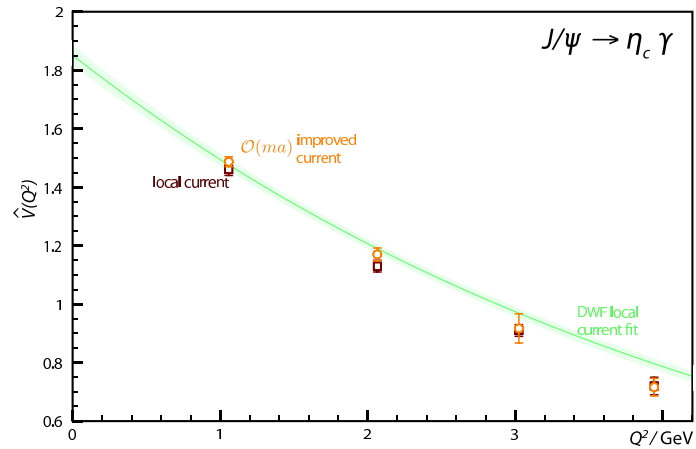}
\vspace{-8mm}
\caption{\label{PVimp}$J/\psi \to \eta_c \gamma$ transition form-factor using local and ${\cal O}(ma)$ improved current with anisotropic Clover quarks and the fit from a study using the local current with DWF fermions on the same anisotropic lattices.}
\end{figure}

We computed form-factors using the improved currents for a few of the transitions considered in this paper. The only effect of a considerable size was found in the $\chi_{c0} \to J/\psi \gamma$ transition as shown in figure \ref{SVimp}(a). Note that the addition of the improvement brings the Clover data into better agreement with the DWF data on the same lattices as we might expect given the automatic ${\cal O}(a)$ improvement one has with DWF. We note that for the same correlators projected on to the excited state $\psi'$ such a large difference with respect to the local current was not seen, see figure \ref{SVimp}(b). We were unable to find any other large effects due to improvement, e.g. consider $J/\psi \to \eta_c \gamma$ shown in figure \ref{PVimp}. These observations may suggest that except in certain notable cases (the scalar), the discretization errors on our results are relatively small.

\bibliography{charmonium_lattice}

\begin{thebibliography}{34}
\expandafter\ifx\csname natexlab\endcsname\relax\def\natexlab#1{#1}\fi
\expandafter\ifx\csname bibnamefont\endcsname\relax
  \def\bibnamefont#1{#1}\fi
\expandafter\ifx\csname bibfnamefont\endcsname\relax
  \def\bibfnamefont#1{#1}\fi
\expandafter\ifx\csname citenamefont\endcsname\relax
  \def\citenamefont#1{#1}\fi
\expandafter\ifx\csname url\endcsname\relax
  \def\url#1{\texttt{#1}}\fi
\expandafter\ifx\csname urlprefix\endcsname\relax\def\urlprefix{URL }\fi
\providecommand{\bibinfo}[2]{#2}
\providecommand{\eprint}[2][]{\url{#2}}

\bibitem[{\citenamefont{Dudek et~al.}(2008)\citenamefont{Dudek, Edwards,
  Mathur, and Richards}}]{Dudek:2007wv}
\bibinfo{author}{\bibfnamefont{J.~J.} \bibnamefont{Dudek}},
  \bibinfo{author}{\bibfnamefont{R.~G.} \bibnamefont{Edwards}},
  \bibinfo{author}{\bibfnamefont{N.}~\bibnamefont{Mathur}}, \bibnamefont{and}
  \bibinfo{author}{\bibfnamefont{D.~G.} \bibnamefont{Richards}},
  \bibinfo{journal}{Phys. Rev.} \textbf{\bibinfo{volume}{D77}},
  \bibinfo{pages}{034501} (\bibinfo{year}{2008}), \eprint{0707.4162}.

\bibitem[{\citenamefont{Dudek et~al.}(2006)\citenamefont{Dudek, Edwards, and
  Richards}}]{Dudek:2006ej}
\bibinfo{author}{\bibfnamefont{J.~J.} \bibnamefont{Dudek}},
  \bibinfo{author}{\bibfnamefont{R.~G.} \bibnamefont{Edwards}},
  \bibnamefont{and} \bibinfo{author}{\bibfnamefont{D.~G.}
  \bibnamefont{Richards}}, \bibinfo{journal}{Phys. Rev.}
  \textbf{\bibinfo{volume}{D73}}, \bibinfo{pages}{074507}
  (\bibinfo{year}{2006}), \eprint{hep-ph/0601137}.

\bibitem[{\citenamefont{Gaiser et~al.}(1986)}]{Gaiser:1985ix}
\bibinfo{author}{\bibfnamefont{J.}~\bibnamefont{Gaiser}} \bibnamefont{et~al.},
  \bibinfo{journal}{Phys. Rev.} \textbf{\bibinfo{volume}{D34}},
  \bibinfo{pages}{711} (\bibinfo{year}{1986}).

\bibitem[{\citenamefont{Mitchell et~al.}(2009)}]{Mitchell:2008fb}
\bibinfo{author}{\bibfnamefont{R.~E.} \bibnamefont{Mitchell}}
  \bibnamefont{et~al.} (\bibinfo{collaboration}{CLEO}), \bibinfo{journal}{Phys.
  Rev. Lett.} \textbf{\bibinfo{volume}{102}}, \bibinfo{pages}{011801}
  (\bibinfo{year}{2009}), \eprint{0805.0252}.

\bibitem[{\citenamefont{Barnes and Swanson}(2008)}]{Barnes:2007xu}
\bibinfo{author}{\bibfnamefont{T.}~\bibnamefont{Barnes}} \bibnamefont{and}
  \bibinfo{author}{\bibfnamefont{E.~S.} \bibnamefont{Swanson}},
  \bibinfo{journal}{Phys. Rev.} \textbf{\bibinfo{volume}{C77}},
  \bibinfo{pages}{055206} (\bibinfo{year}{2008}), \eprint{0711.2080}.

\bibitem[{\citenamefont{Eichten et~al.}(1978)\citenamefont{Eichten, Gottfried,
  Kinoshita, Lane, and Yan}}]{Eichten:1978tg}
\bibinfo{author}{\bibfnamefont{E.}~\bibnamefont{Eichten}},
  \bibinfo{author}{\bibfnamefont{K.}~\bibnamefont{Gottfried}},
  \bibinfo{author}{\bibfnamefont{T.}~\bibnamefont{Kinoshita}},
  \bibinfo{author}{\bibfnamefont{K.~D.} \bibnamefont{Lane}}, \bibnamefont{and}
  \bibinfo{author}{\bibfnamefont{T.-M.} \bibnamefont{Yan}},
  \bibinfo{journal}{Phys. Rev.} \textbf{\bibinfo{volume}{D17}},
  \bibinfo{pages}{3090} (\bibinfo{year}{1978}).

\bibitem[{\citenamefont{Eichten et~al.}(1980)\citenamefont{Eichten, Gottfried,
  Kinoshita, Lane, and Yan}}]{Eichten:1979ms}
\bibinfo{author}{\bibfnamefont{E.}~\bibnamefont{Eichten}},
  \bibinfo{author}{\bibfnamefont{K.}~\bibnamefont{Gottfried}},
  \bibinfo{author}{\bibfnamefont{T.}~\bibnamefont{Kinoshita}},
  \bibinfo{author}{\bibfnamefont{K.~D.} \bibnamefont{Lane}}, \bibnamefont{and}
  \bibinfo{author}{\bibfnamefont{T.-M.} \bibnamefont{Yan}},
  \bibinfo{journal}{Phys. Rev.} \textbf{\bibinfo{volume}{D21}},
  \bibinfo{pages}{203} (\bibinfo{year}{1980}).

\bibitem[{\citenamefont{Li and Zhao}(2008)}]{Li:2007xr}
\bibinfo{author}{\bibfnamefont{G.}~\bibnamefont{Li}} \bibnamefont{and}
  \bibinfo{author}{\bibfnamefont{Q.}~\bibnamefont{Zhao}},
  \bibinfo{journal}{Phys. Lett.} \textbf{\bibinfo{volume}{B670}},
  \bibinfo{pages}{55} (\bibinfo{year}{2008}), \eprint{0709.4639}.

\bibitem[{\citenamefont{Meyer}(2006)}]{Meyer:2006az}
\bibinfo{author}{\bibfnamefont{C.~A.} \bibnamefont{Meyer}},
  \bibinfo{journal}{AIP Conf. Proc.} \textbf{\bibinfo{volume}{870}},
  \bibinfo{pages}{390} (\bibinfo{year}{2006}).

\bibitem[{\citenamefont{Close and Dudek}(2003)}]{Close:2003fz}
\bibinfo{author}{\bibfnamefont{F.~E.} \bibnamefont{Close}} \bibnamefont{and}
  \bibinfo{author}{\bibfnamefont{J.~J.} \bibnamefont{Dudek}},
  \bibinfo{journal}{Phys. Rev. Lett.} \textbf{\bibinfo{volume}{91}},
  \bibinfo{pages}{142001} (\bibinfo{year}{2003}), \eprint{hep-ph/0304243}.

\bibitem[{\citenamefont{Close and Dudek}(2004)}]{Close:2003ae}
\bibinfo{author}{\bibfnamefont{F.~E.} \bibnamefont{Close}} \bibnamefont{and}
  \bibinfo{author}{\bibfnamefont{J.~J.} \bibnamefont{Dudek}},
  \bibinfo{journal}{Phys. Rev.} \textbf{\bibinfo{volume}{D69}},
  \bibinfo{pages}{034010} (\bibinfo{year}{2004}), \eprint{hep-ph/0308098}.

\bibitem[{\citenamefont{Dudek and Rrapaj}(2008)}]{Dudek:2008sz}
\bibinfo{author}{\bibfnamefont{J.~J.} \bibnamefont{Dudek}} \bibnamefont{and}
  \bibinfo{author}{\bibfnamefont{E.}~\bibnamefont{Rrapaj}},
  \bibinfo{journal}{Phys. Rev.} \textbf{\bibinfo{volume}{D78}},
  \bibinfo{pages}{094504} (\bibinfo{year}{2008}), \eprint{0809.2582}.

\bibitem[{\citenamefont{Burch et~al.}(2009)\citenamefont{Burch, Hagen, Lang,
  Limmer, and Schafer}}]{Burch:2008qx}
\bibinfo{author}{\bibfnamefont{T.}~\bibnamefont{Burch}},
  \bibinfo{author}{\bibfnamefont{C.}~\bibnamefont{Hagen}},
  \bibinfo{author}{\bibfnamefont{C.~B.} \bibnamefont{Lang}},
  \bibinfo{author}{\bibfnamefont{M.}~\bibnamefont{Limmer}}, \bibnamefont{and}
  \bibinfo{author}{\bibfnamefont{A.}~\bibnamefont{Schafer}},
  \bibinfo{journal}{Phys. Rev.} \textbf{\bibinfo{volume}{D79}},
  \bibinfo{pages}{014504} (\bibinfo{year}{2009}), \eprint{0809.1103}.

\bibitem[{\citenamefont{Blossier et~al.}(2008)\citenamefont{Blossier, von
  Hippel, Mendes, Sommer, and Della~Morte}}]{Blossier:2008tx}
\bibinfo{author}{\bibfnamefont{B.}~\bibnamefont{Blossier}},
  \bibinfo{author}{\bibfnamefont{G.}~\bibnamefont{von Hippel}},
  \bibinfo{author}{\bibfnamefont{T.}~\bibnamefont{Mendes}},
  \bibinfo{author}{\bibfnamefont{R.}~\bibnamefont{Sommer}}, \bibnamefont{and}
  \bibinfo{author}{\bibfnamefont{M.}~\bibnamefont{Della~Morte}},
  \bibinfo{journal}{PoS} \textbf{\bibinfo{volume}{LATTICE2008}},
  \bibinfo{pages}{135} (\bibinfo{year}{2008}), \eprint{0808.1017}.

\bibitem[{\citenamefont{Amsler et~al.}(2008)}]{Amsler:2008zzb}
\bibinfo{author}{\bibfnamefont{C.}~\bibnamefont{Amsler}} \bibnamefont{et~al.}
  (\bibinfo{collaboration}{Particle Data Group}), \bibinfo{journal}{Phys.
  Lett.} \textbf{\bibinfo{volume}{B667}}, \bibinfo{pages}{1}
  (\bibinfo{year}{2008}).

\bibitem[{\citenamefont{Briere et~al.}(2006)}]{Briere:2006ff}
\bibinfo{author}{\bibfnamefont{R.~A.} \bibnamefont{Briere}}
  \bibnamefont{et~al.} (\bibinfo{collaboration}{CLEO}), \bibinfo{journal}{Phys.
  Rev.} \textbf{\bibinfo{volume}{D74}}, \bibinfo{pages}{031106}
  (\bibinfo{year}{2006}), \eprint{hep-ex/0605070}.

\bibitem[{\citenamefont{Bardeen et~al.}(2001)\citenamefont{Bardeen, Duncan,
  Eichten, Isgur, and Thacker}}]{Bardeen:2001jm}
\bibinfo{author}{\bibfnamefont{W.~A.} \bibnamefont{Bardeen}},
  \bibinfo{author}{\bibfnamefont{A.}~\bibnamefont{Duncan}},
  \bibinfo{author}{\bibfnamefont{E.}~\bibnamefont{Eichten}},
  \bibinfo{author}{\bibfnamefont{N.}~\bibnamefont{Isgur}}, \bibnamefont{and}
  \bibinfo{author}{\bibfnamefont{H.}~\bibnamefont{Thacker}},
  \bibinfo{journal}{Phys. Rev.} \textbf{\bibinfo{volume}{D65}},
  \bibinfo{pages}{014509} (\bibinfo{year}{2001}), \eprint{hep-lat/0106008}.

\bibitem[{\citenamefont{Barnes et~al.}(2005)\citenamefont{Barnes, Godfrey, and
  Swanson}}]{Barnes:2005pb}
\bibinfo{author}{\bibfnamefont{T.}~\bibnamefont{Barnes}},
  \bibinfo{author}{\bibfnamefont{S.}~\bibnamefont{Godfrey}}, \bibnamefont{and}
  \bibinfo{author}{\bibfnamefont{E.~S.} \bibnamefont{Swanson}},
  \bibinfo{journal}{Phys. Rev.} \textbf{\bibinfo{volume}{D72}},
  \bibinfo{pages}{054026} (\bibinfo{year}{2005}), \eprint{hep-ph/0505002}.

\bibitem[{\citenamefont{Godfrey and Isgur}(1985)}]{Godfrey:1985xj}
\bibinfo{author}{\bibfnamefont{S.}~\bibnamefont{Godfrey}} \bibnamefont{and}
  \bibinfo{author}{\bibfnamefont{N.}~\bibnamefont{Isgur}},
  \bibinfo{journal}{Phys. Rev.} \textbf{\bibinfo{volume}{D32}},
  \bibinfo{pages}{189} (\bibinfo{year}{1985}).

\bibitem[{\citenamefont{Isgur and Paton}(1985)}]{Isgur:1984bm}
\bibinfo{author}{\bibfnamefont{N.}~\bibnamefont{Isgur}} \bibnamefont{and}
  \bibinfo{author}{\bibfnamefont{J.~E.} \bibnamefont{Paton}},
  \bibinfo{journal}{Phys. Rev.} \textbf{\bibinfo{volume}{D31}},
  \bibinfo{pages}{2910} (\bibinfo{year}{1985}).

\bibitem[{\citenamefont{Guo et~al.}(2008)\citenamefont{Guo, Szczepaniak,
  Galata, Vassallo, and Santopinto}}]{Guo:2008yz}
\bibinfo{author}{\bibfnamefont{P.}~\bibnamefont{Guo}},
  \bibinfo{author}{\bibfnamefont{A.~P.} \bibnamefont{Szczepaniak}},
  \bibinfo{author}{\bibfnamefont{G.}~\bibnamefont{Galata}},
  \bibinfo{author}{\bibfnamefont{A.}~\bibnamefont{Vassallo}}, \bibnamefont{and}
  \bibinfo{author}{\bibfnamefont{E.}~\bibnamefont{Santopinto}},
  \bibinfo{journal}{Phys. Rev.} \textbf{\bibinfo{volume}{D78}},
  \bibinfo{pages}{056003} (\bibinfo{year}{2008}), \eprint{0807.2721}.

\bibitem[{\citenamefont{Eichten et~al.}(2004)\citenamefont{Eichten, Lane, and
  Quigg}}]{Eichten:2004uh}
\bibinfo{author}{\bibfnamefont{E.~J.} \bibnamefont{Eichten}},
  \bibinfo{author}{\bibfnamefont{K.}~\bibnamefont{Lane}}, \bibnamefont{and}
  \bibinfo{author}{\bibfnamefont{C.}~\bibnamefont{Quigg}},
  \bibinfo{journal}{Phys. Rev.} \textbf{\bibinfo{volume}{D69}},
  \bibinfo{pages}{094019} (\bibinfo{year}{2004}), \eprint{hep-ph/0401210}.

\bibitem[{\citenamefont{Chen}(2001)}]{Chen:2000ej}
\bibinfo{author}{\bibfnamefont{P.}~\bibnamefont{Chen}}, \bibinfo{journal}{Phys.
  Rev.} \textbf{\bibinfo{volume}{D64}}, \bibinfo{pages}{034509}
  (\bibinfo{year}{2001}), \eprint{hep-lat/0006019}.

\bibitem[{\citenamefont{Harada et~al.}(2001)\citenamefont{Harada, Kronfeld,
  Matsufuru, Nakajima, and Onogi}}]{Harada:2001ei}
\bibinfo{author}{\bibfnamefont{J.}~\bibnamefont{Harada}},
  \bibinfo{author}{\bibfnamefont{A.~S.} \bibnamefont{Kronfeld}},
  \bibinfo{author}{\bibfnamefont{H.}~\bibnamefont{Matsufuru}},
  \bibinfo{author}{\bibfnamefont{N.}~\bibnamefont{Nakajima}}, \bibnamefont{and}
  \bibinfo{author}{\bibfnamefont{T.}~\bibnamefont{Onogi}},
  \bibinfo{journal}{Phys. Rev.} \textbf{\bibinfo{volume}{D64}},
  \bibinfo{pages}{074501} (\bibinfo{year}{2001}), \eprint{hep-lat/0103026}.

\bibitem[{\citenamefont{Boyle et~al.}(2007)\citenamefont{Boyle, Flynn, Juttner,
  Sachrajda, and Zanotti}}]{Boyle:2007wg}
\bibinfo{author}{\bibfnamefont{P.~A.} \bibnamefont{Boyle}},
  \bibinfo{author}{\bibfnamefont{J.~M.} \bibnamefont{Flynn}},
  \bibinfo{author}{\bibfnamefont{A.}~\bibnamefont{Juttner}},
  \bibinfo{author}{\bibfnamefont{C.~T.} \bibnamefont{Sachrajda}},
  \bibnamefont{and} \bibinfo{author}{\bibfnamefont{J.~M.}
  \bibnamefont{Zanotti}}, \bibinfo{journal}{JHEP}
  \textbf{\bibinfo{volume}{05}}, \bibinfo{pages}{016} (\bibinfo{year}{2007}),
  \eprint{hep-lat/0703005}.

\bibitem[{\citenamefont{Lakhina and Swanson}(2006)}]{Lakhina:2006vg}
\bibinfo{author}{\bibfnamefont{O.}~\bibnamefont{Lakhina}} \bibnamefont{and}
  \bibinfo{author}{\bibfnamefont{E.~S.} \bibnamefont{Swanson}},
  \bibinfo{journal}{Phys. Rev.} \textbf{\bibinfo{volume}{D74}},
  \bibinfo{pages}{014012} (\bibinfo{year}{2006}), \eprint{hep-ph/0603164}.

\bibitem[{\citenamefont{Olsson et~al.}(1985)\citenamefont{Olsson, Suchyta,
  Martin, and Stirling}}]{Olsson:1984zm}
\bibinfo{author}{\bibfnamefont{M.~G.} \bibnamefont{Olsson}},
  \bibinfo{author}{\bibfnamefont{I.}~\bibnamefont{Suchyta},
  \bibfnamefont{C.~J.}}, \bibinfo{author}{\bibfnamefont{A.~D.}
  \bibnamefont{Martin}}, \bibnamefont{and}
  \bibinfo{author}{\bibfnamefont{W.~J.} \bibnamefont{Stirling}},
  \bibinfo{journal}{Phys. Rev.} \textbf{\bibinfo{volume}{D31}},
  \bibinfo{pages}{1759} (\bibinfo{year}{1985}).

\bibitem[{\citenamefont{Eichten et~al.}(2002)\citenamefont{Eichten, Lane, and
  Quigg}}]{Eichten:2002qv}
\bibinfo{author}{\bibfnamefont{E.~J.} \bibnamefont{Eichten}},
  \bibinfo{author}{\bibfnamefont{K.}~\bibnamefont{Lane}}, \bibnamefont{and}
  \bibinfo{author}{\bibfnamefont{C.}~\bibnamefont{Quigg}},
  \bibinfo{journal}{Phys. Rev. Lett.} \textbf{\bibinfo{volume}{89}},
  \bibinfo{pages}{162002} (\bibinfo{year}{2002}), \eprint{hep-ph/0206018}.

\bibitem[{\citenamefont{Brambilla et~al.}(2006)\citenamefont{Brambilla, Jia,
  and Vairo}}]{Brambilla:2005zw}
\bibinfo{author}{\bibfnamefont{N.}~\bibnamefont{Brambilla}},
  \bibinfo{author}{\bibfnamefont{Y.}~\bibnamefont{Jia}}, \bibnamefont{and}
  \bibinfo{author}{\bibfnamefont{A.}~\bibnamefont{Vairo}},
  \bibinfo{journal}{Phys. Rev.} \textbf{\bibinfo{volume}{D73}},
  \bibinfo{pages}{054005} (\bibinfo{year}{2006}), \eprint{hep-ph/0512369}.

\bibitem[{\citenamefont{Beilin and Radyushkin}(1984)}]{Beilin:1983ht}
\bibinfo{author}{\bibfnamefont{V.~A.} \bibnamefont{Beilin}} \bibnamefont{and}
  \bibinfo{author}{\bibfnamefont{A.~V.} \bibnamefont{Radyushkin}},
  \bibinfo{journal}{Sov. J. Nucl. Phys.} \textbf{\bibinfo{volume}{39}},
  \bibinfo{pages}{800} (\bibinfo{year}{1984}).

\bibitem[{\citenamefont{Close and Swanson}(2005)}]{Close:2005se}
\bibinfo{author}{\bibfnamefont{F.~E.} \bibnamefont{Close}} \bibnamefont{and}
  \bibinfo{author}{\bibfnamefont{E.~S.} \bibnamefont{Swanson}},
  \bibinfo{journal}{Phys. Rev.} \textbf{\bibinfo{volume}{D72}},
  \bibinfo{pages}{094004} (\bibinfo{year}{2005}), \eprint{hep-ph/0505206}.

\bibitem[{\citenamefont{Close and Thomas}(2009)}]{Close:2009ii}
\bibinfo{author}{\bibfnamefont{F.~E.} \bibnamefont{Close}} \bibnamefont{and}
  \bibinfo{author}{\bibfnamefont{C.~E.} \bibnamefont{Thomas}}
  (\bibinfo{year}{2009}), \eprint{0901.1812}.

\bibitem[{\citenamefont{Uehara et~al.}(2006)}]{Uehara:2005qd}
\bibinfo{author}{\bibfnamefont{S.}~\bibnamefont{Uehara}} \bibnamefont{et~al.}
  (\bibinfo{collaboration}{Belle}), \bibinfo{journal}{Phys. Rev. Lett.}
  \textbf{\bibinfo{volume}{96}}, \bibinfo{pages}{082003}
  (\bibinfo{year}{2006}), \eprint{hep-ex/0512035}.

\bibitem[{\citenamefont{Dudek and Edwards}(2006)}]{Dudek:2006ut}
\bibinfo{author}{\bibfnamefont{J.~J.} \bibnamefont{Dudek}} \bibnamefont{and}
  \bibinfo{author}{\bibfnamefont{R.~G.} \bibnamefont{Edwards}},
  \bibinfo{journal}{Phys. Rev. Lett.} \textbf{\bibinfo{volume}{97}},
  \bibinfo{pages}{172001} (\bibinfo{year}{2006}), \eprint{hep-ph/0607140}.

\end{thebibliography}

\end{document}